\documentclass[%
superscriptaddress,
 amsmath,
 amssymb,
 aps,
 prl,
floatfix,
twocolumn
]{revtex4-1}
\usepackage[utf8]{inputenc}
\usepackage[T1]{fontenc}
\usepackage{babel}
\usepackage{graphicx}
\usepackage{amsfonts}
\usepackage{amssymb}
\usepackage{amsmath}
\usepackage{amsthm}
\usepackage{mathtools}
\usepackage{physics}
\usepackage[normalem]{ulem}
\usepackage{blindtext}
\usepackage{dsfont}

\usepackage{mathtools}
\usepackage{braket}
\renewcommand\bra[1]{{\langle{#1}|}}
\renewcommand\ket[1]{{|{#1}\rangle}}
\usepackage{graphicx}
\usepackage[format=plain,justification=centerlast]{caption}
\usepackage[format=plain,justification=centerlast]{subcaption}
\usepackage{bm}
\usepackage[dvipsnames]{xcolor}

\usepackage{natbib}
\setcitestyle{square,numbers}
\usepackage[colorlinks]{hyperref}
\usepackage{changes}
\definechangesauthor[name=Giacomo, color=red]{GS}
\definechangesauthor[name=Tomasz, color=ForestGreen]{TL}
\definechangesauthor[name=Konrad, color=Magenta]{KS}

\begin{document}

\title{Practical tests for sub-Rayleigh source discriminations with imperfect demultiplexers} 
\author{Konrad Schlichtholz}
\affiliation{International Centre for Theory of Quantum Technologies, University of Gdansk, 80-308 Gda{\'n}sk, Poland}

\author{Tomasz Linowski}
\affiliation{International Centre for Theory of Quantum Technologies, University of Gdansk, 80-308 Gda{\'n}sk, Poland}

\author{Mattia Walschaers}
\affiliation{Laboratoire Kastler Brossel, Sorbonne Université, ENS-Université PSL, CNRS, Collège de France, 4 Place Jussieu, 75252 Paris, France}

\author{Nicolas Treps}
\affiliation{Laboratoire Kastler Brossel, Sorbonne Université, ENS-Université PSL, CNRS, Collège de France, 4 Place Jussieu, 75252 Paris, France}

\author{{\L}ukasz Rudnicki}
\affiliation{International Centre for Theory of Quantum Technologies, University of Gdansk, 80-308 Gda{\'n}sk, Poland}
\affiliation{Center for Theoretical Physics, Polish Academy of Sciences, 02-668 Warszawa, Poland}
\author{Giacomo Sorelli}
\affiliation{Laboratoire Kastler Brossel, Sorbonne Université, ENS-Université PSL, CNRS, Collège de France, 4 Place Jussieu, 75252 Paris, France}
\affiliation{Fraunhofer IOSB, Ettlingen, Fraunhofer Institute of Optronics,
System Technologies and Image Exploitation, Gutleuthausstr. 1, 76275 Ettlingen, Germany}  

\pagenumbering{arabic}

\begin{abstract}
Quantum-optimal discrimination between one and two closely separated light sources can be achieved by ideal spatial-mode demultiplexing, simply monitoring whether a photon is detected in a single antisymmetric mode. However, we show that for any, no matter how small, imperfections of the demultiplexer, this simple statistical test becomes practically useless, i.e. as good as flipping a coin. While we identify a class of separation-independent tests with vanishing error probabilities in the limit of large numbers of detected photons, they are generally unreliable beyond that very limit. As a practical alternative, we propose a simple semi-separation-independent test, which provides a method for designing reliable experiments, through arbitrary control over the maximal probability of error.

\end{abstract}

\maketitle

\paragraph{Introduction --}
Statistical hypothesis testing is an important tool in the analysis of scientific data. 
A typical hypothesis testing problem in optical imaging is source discrimination, i.e. establishing whether an image originates from one or two light sources \cite{Harris:64}. This is relevant in astronomy, e.g. for efficient exoplanet and binary stars detection \cite{Exo_1,Exo_2,Exo_3,Exo_7,Exo_6,Exo_4,Exo_5}, and fluorescence microscopy, e.g. for counting the exact number of molecules in a sample \cite{Micro_2,Micro_1,Micro_3}.
For source separations smaller than the width of the point spread function of the optical apparatus, the efficiency of source discrimination protocols based on spatially resolved intensity measurements, i.e.  direct imaging, drops significantly \cite{resolution_survey_denDekker_1997,Fourier_optics_Goodman_2005}. 
In this \emph{sub-Rayleigh} regime, source discrimination could be performed with the help of quantum-inspired measurement techniques \cite{Helstrom}.

Recently,  motivated  by the super-resolving power of spatial demultiplexing (SPADE) in the closely related problem of source separation estimation \cite{superresolution_Tsang_2016,superresolution_starlight_Tsang_2019}, it was shown that SPADE is also quantum-optimal in source discrimination \cite{HYP_1}, even when the sources are not point-like \cite{no_point}. 
Furthermore, it was demonstrated that, due to the symmetry of the problem, detecting even just one photon in a fixed antisymmetric mode allows to accept one of the hypotheses with zero probability of error, leading to a near-optimal, separation-independent decision strategy  \cite{HYP_1}. 
These findings, however, were obtained assuming  ideal measurements. In practice, the obtained results are significantly affected by experimental imperfections \cite{noon_advances,metrology_noise_Kolodynski_2013,metrology_noise_Nichols_2016,noise_Oh_2021,Sorelli_practical_superresolution_2021,imaging_noisy_Lupo_2020,superresolution_limits_SPADE_Len_2020,resolution_misalingment_Almeida_2021,superresolution_no_location_Grace_2020}. In particular, in the case of SPADE, misalignment, defects in the fabrication of the demultiplexer and other imperfections induce a finite probability of detecting photons in the incorrect output, i.e. \emph{crosstalk} \cite{Manuel,Giacomo,Linowski}.

In this Letter, we show that crosstalk has a strong impact on SPADE for discriminating between one and two equally bright sources in the practically relevant regime of small separations. In particular, we prove that the simple separation-independent test discussed above changes from being quantum-optimal in the ideal case, to be as good as flipping a coin in presence of arbitrarily small crosstalk. Moreover, we find that even though is possible to design a class of meaningful separation-independent tests even in presence of crosstalk, the associated error probabilities are hard to predict without previous knowledge of the source separations. As an alternative, we propose a semi-separation-independent test with easily accessible maximal probability of error.

\begin{figure}[t]
    \centering
         \centering
         \includegraphics[width=0.49\textwidth]{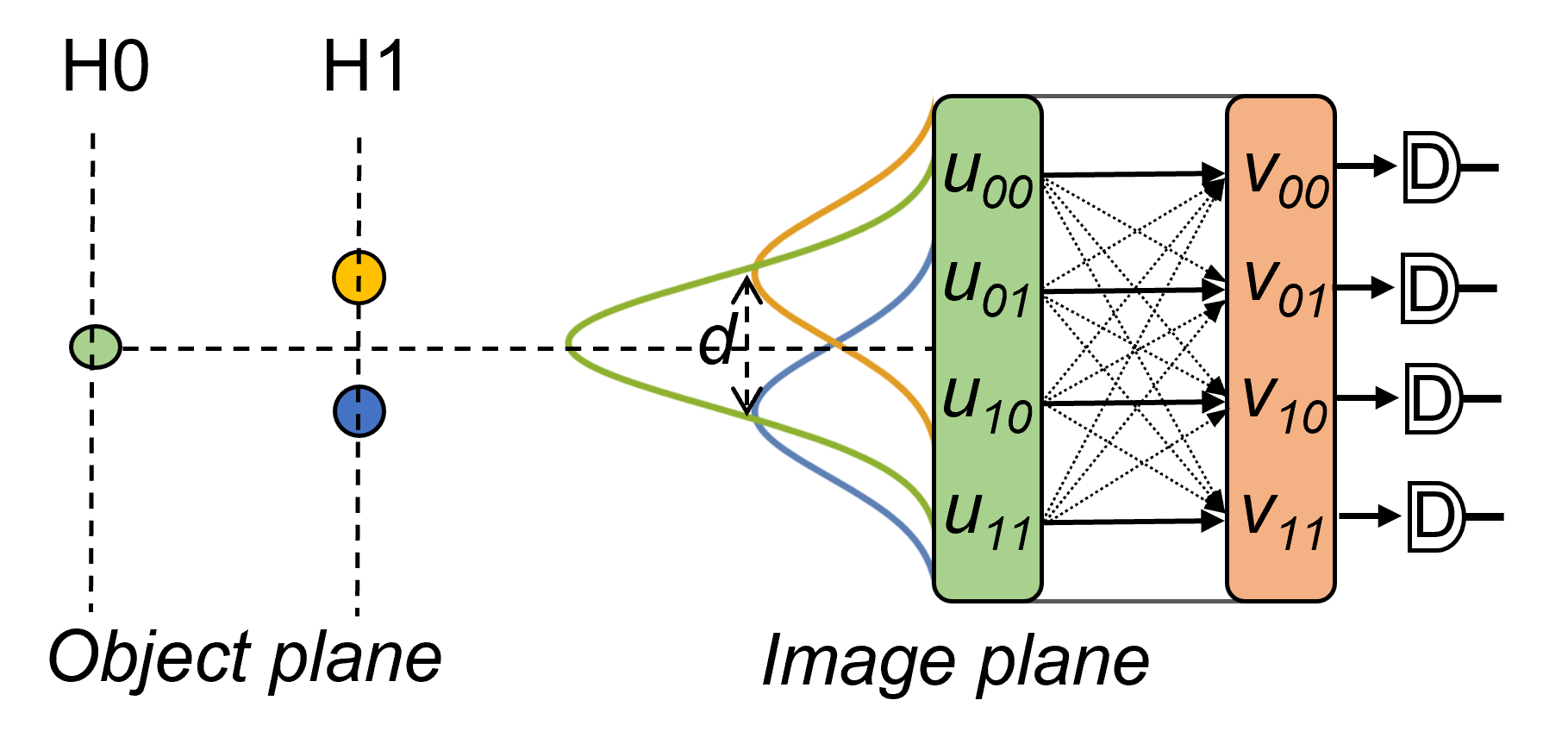}
        \caption{Schematic representation of the measurement scenario. Depending on the hypothesis, there is one (H0) or two (H1) weak light sources in the object plane, resulting in diffraction-broadened spatial field distributions in the image plane. To decide whether H0 or H1 is true the image-plane field distribution is analyzed via photon counting after spatial-mode demultiplexing affected by crosstalk.}
        \label{fig:system}  
\end{figure}

\paragraph{Hypotheses and measurement setting --}
We are interested in distinguishing between two hypotheses, H0 and H1, as illustrated in Fig. \ref{fig:system}. According to hypothesis H1, two weak, incoherent light sources of equal brightness (e.g. faraway thermal sources) are separated by a distance $d$ in the object plane. 
The coordinate system is chosen in such a way that the sources' positions are given by $\pm\vec{r}_0=\pm (d/2, 0)$. 
We consider a diffraction-limited imaging system with a Gaussian point spread function $u_{00}(\vec{r})=\sqrt{2/(\pi w^2)}\exp{-r^2/w^2}$, so that the spatial distribution of the electromagnetic field in the image plane coming from a source at $\pm\vec{r}_0$ is given by $u_{00}(\vec{r}\mp\vec{r}_0)$ \cite{goodman1985}.
For weak sources,  most of the photon detection events are single-photon events. Therefore,  results  in the image plane are effectively described as repeated measurements on $N$ copies of the single-photon state \cite{superresolution_Tsang_2016}
\begin{equation}
    \hat{\rho}_{\textnormal{H1}}(d)\approx 
        \frac{1}{2}\Big( \ket{\phi({d})}\bra{\phi({d})}
        +\ket{\phi(-{d})}\bra{\phi(-{d})}\Big),\label{eq:state}
\end{equation}
where  
$ \ket{\phi(\pm{d})}=\int d\vec{r}\;u_{00}(\vec{r}\mp\vec{r}_0)\ket{\vec{r}} $
and $\ket{\vec{r}}$ stands for the single-photon position eigenstate in the image plane. 
According to hypothesis H0, there is only one source in the object plane, centered at the origin of the coordinate system, and with the same total brightness as the two sources from hypothesis H1. 
In this case, the measurement results are effectively described by repeated measurements on the state
\begin{equation}
    \hat{\rho}_{\textnormal{H0}} = \lim_{d\to 0}\hat{\rho}_{\textnormal{H1}}(d)
        =\ket{\phi(0)}\bra{\phi(0)}.
\end{equation}

The efficiency of a given strategy for deciding which one of the two hypotheses is true is captured by the average probability of error 
\begin{equation}
    P_{\textnormal{e}}(N)=P_{\textnormal{H0}}\alpha(N)+P_{\textnormal{H1}}\beta(N),
\end{equation}
where $P_{\textnormal{H0(1)}}$ are the a priori probabilities for the respective hypothesis and $\alpha(N)$, $\beta(N)$ are the probabilities of error of the first and second kind, i.e. assuming H1 when H0 is correct and vice versa, for a sample of size $N$. If no a priori information is available  there is no reason to assign higher probability to any of the hypotheses. For clarity, we concentrate on such a case in which $P_{\textnormal{H0}}=P_{\textnormal{H1}}=1/2$. Note that, abandoning this assumption has no qualitative impact on our results.

\paragraph{Asymptotic probability of error --}
Effective source discrimination for sub-Rayleigh separations, i.e. for $x:=d/2w<1$, requires a large number $N$ of samples. 
When $N\gg1$, the  probability of error minimized over all possible decision strategies for a specific measurement decays exponentially as $P_{\textnormal{e}}^{\min}\sim\exp{-N\xi}$, where
\begin{align}
    \xi=-\ln(\min_{0\leq s\leq1}\sum_{k}p(k|\textnormal{H0})^s
        p(k|\textnormal{H1})^{1-s}).\label{eq:chernoff}
\end{align}
is the \emph{Chernoff exponent} \cite{Chernoff}. Here, $p(k|\textnormal{H0})$ ($p(k|\textnormal{H1})$)stand for the probability of obtaining the measurement outcome $k$ conditioned on the  hypothesis H0 (H1)  being true. 
By optimizing Eq.~\eqref{eq:chernoff} over all possible measurements, we obtain the \emph{quantum Chernoff exponent}  \cite{Q_Chernoff}. 
It was shown that for the considered hypothesis testing problem (with weak incoherent sources), the quantum Chernoff exponent is given by $\xi_Q=x^2$ and that it can be saturated by spatial demultiplexing (SPADE) in Hermite-Gauss modes  $u_{nm}$ centered at the origin of our coordinate system  \cite{HYP_1}.

However, in  experimental settings, any measurement is unavoidably subject to apparatus misalignment, design and fabrication defects of the demultiplexer, and other imperfections.  Accordingly, there is a small crosstalk probability, i.e. it is possible that a measured photon is transmitted to an incorrect mode (see Fig.~\ref{fig:system}).
Consequently, the real measurement basis deviates from the ideal one as 
$
    v_{nm} = \sum_{k,l=0}^{D-1} C_{nm,kl}u_{kl},
$
where $C$ stands for the (unitary) crosstalk matrix and $D$ restricts the number of measured modes \cite{appendix}. For small separations, the optimal probability of error is achieved already with $D=2$, which is what we will assume from now on \cite{HYP_1}. 
To quantify the severity of imperfections, we use crosstalk strength $\epsilon^2$, defined as the mean absolute value square of the off-diagonal  elements of the crosstalk matrix \cite{Manuel,appendix}. 

\begin{figure}[t]
    \centering
         \centering
         \includegraphics[width=0.49\textwidth]{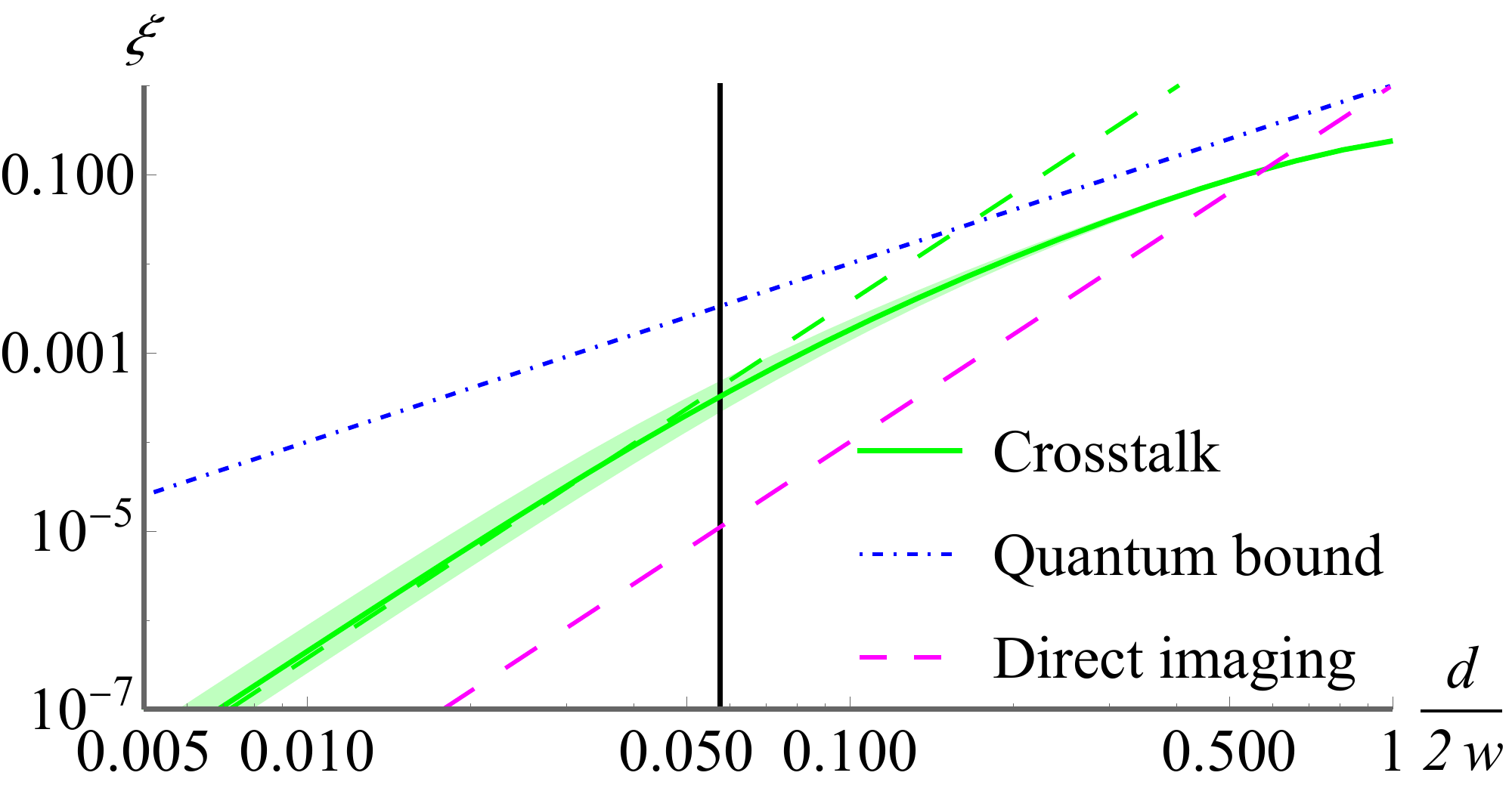}
        \caption{Comparison between the median of the Chernoff exponents for SPADE $\xi$ from a sample of $500$ random unitary crosstalk matrix (green, shaded area stands for the interquartile range), approximate Chernoff exponent \eqref{eq:xi_series} for $x\ll\epsilon$ (green, dashed), asymptotic Chernoff exponent for perfect direct imaging $\xi_{\rm DI}$ (pink, dashed) and the quantum bound $\xi_{\rm Q}$ (blue, dot-dashed) versus $x := d/2w$. Calculations performed with  $\epsilon^2=0.0033$ and $D=2$. The black vertical line indicates $x=\epsilon$. }
        \label{fig:CHernoff}  
\end{figure}

We now proceed to assess the impact of crosstalk on SPADE. 
We focus on the practically relevant regime of small separations, $x\ll1$, and weak crosstalk $\epsilon\ll1$. Accordingly, we can approximate the Chernoff exponent (\ref{eq:chernoff}) by a series expansion in these parameters. Analogously to previous findings for separation estimation  \cite{Linowski}, we find that the expansion depends on the ratio $x/\epsilon$ \cite{appendix}:
\begin{align} 
    \xi\approx
    \begin{cases}
        \left\{1 - \left[\ln\ln q(x)-1\right]/\ln q(x) \right\} x^2 & x\gg \epsilon, \\   
        x^4/(8 p_0), & x\ll \epsilon,
    \end{cases}
    \label{eq:xi_series}
\end{align}
where we restricted ourselves to leading terms in $x$ and $\epsilon$. Here, $q(x)\coloneqq x^2/p_0$ and $p_0\coloneqq |C_{10,00}|^2\sim \epsilon^2$ is the probability of crosstalk from mode $u_{00}$ to $v_{10}$. Note that for $x\approx \epsilon$ the obtained series converges too slowly to constitute a reliable approximation.
In the range $x\ll \epsilon$, crosstalk changes the scaling of $\xi$ from $x^2$ to merely $x^4$. We note that the Chernoff exponent for ideal direct imaging is given by $\xi_{\rm DI} = x^4$ \cite{appendix}. 
Nonetheless, despite the same scaling, SPADE is still superior to direct imaging due to a larger scaling coefficient (for weak crosstalk, $1/(8p_0)\gg 1$). 
More surprisingly, crosstalk has a significant impact on the Chernoff exponent even in the range of relatively large separations $x\gg \epsilon$. 
Indeed, while the upper line of Eq.~\eqref{eq:xi_series} approaches the ideal scaling $x^2$ with vanishing crosstalk, it does so logarithmically slowly. 
This shows the importance of  crosstalk in  hypothesis testing at any separation scale. 
A graphical comparison between the Chernoff exponents for crosstalk-affected SPADE $\xi$, ideal direct imaging $\xi_{\rm DI}$ and the quantum bound $\xi_{\rm Q}$ in the sub-Rayleigh regime is provided in Fig. \ref{fig:CHernoff}.

\paragraph{Practicality of statistical tests --}
The optimal decision strategy for a specific measurement is given by the \emph{likelihood-ratio test} \cite{Van_Trees}, according to which H1 is accepted if and only if
\begin{align} \label{eq:optimal_test}
    \prod_{nm}\left(\frac{p(nm|d,D)}{p(nm|0,D)}\right)^{N_{nm}} > 1,
\end{align}
where $N_{nm}$ is the number of measured photons in mode $v_{nm}$ and $p(nm|d,D)$ is the probability of measuring a photon in this mode for a separation $d$. 
Unfortunately, the likelihood ratio test has some drawbacks. Due to its assumption of a fixed separation, the latter has to be first estimated from the data, using, e.g., the method of moments \cite{Giacomo}. 
Errors in this estimation inevitably deviate this decision strategy from its optimality, and more importantly can lead to underestimation of the probability of error, which is in addition hard to calculate.
Finally, the optimal test \eqref{eq:optimal_test} requires measuring in many modes, which is not always feasible in experiments.

A seemingly more practical test was introduced in Ref.~\cite{HYP_1}. 
The key observation behind this test is that, assuming ideal measurements, $p(10|0)=0$, while $p(10|x)> 0$. 
In other words, photons can be measured in mode $v_{10}$ only if hypothesis H1 is true. 
This leads to the following test:
\begin{align} \label{eq:original_test}
\begin{split}
    N_{10} \underset{H0}{\overset{H1}{\gtrless}} 0,
\end{split}
\end{align}
where whenever equality holds one assumes H0. One can easily calculate that for such test $\alpha(N)=0$ and $\beta(N) = \left[1 - p(10|x)\right]^N$ \cite{HYP_1}. 
Clearly, both $\alpha(N)$ and $\beta(N)$ vanish for large photon numbers $N$ regardless of $x$, and therefore, so does the total probability of error. 
We thus have a test that is simple, separation-independent, and requires measuring in only one mode. 
Moreover, for small separations $x \ll 1$, this simple test is also quantum optimal, as we can show that Eq.~(\ref{eq:original_test}) is in fact equivalent to Eq.~(\ref{eq:optimal_test}) in this regime \cite{appendix}.
Unfortunately, as we will now prove this test goes from optimal to completely useless in presence of any, no matter how small, amount of crosstalk.

To see this, we observe that even for very weak crosstalk $\epsilon \ll 1$, the probability of measuring a photon in mode $v_{10}$ under hypothesis H0 is no longer zero, but rather equals $p(10|0) = p_0 > 0$. More generally, we denote $p_x:=p(10|x)$. 
As long as $p_0$ is non-zero, however small, the asymptotic probability of error is drastically changed. 
It is easily seen that $N_{10}$ for test (\ref{eq:original_test}) is a random variable with binomial distribution, with probabilities given by
\begin{equation}
    p(N_{10}=k|x)=\binom{N}{k} p_x^k\left(1-p_x\right)^{N-k},\label{eq:binom}
\end{equation}
where the corresponding probability for hypothesis H0 is obtained setting $x=0$.
As an immediate consequence, the probability $\alpha(N)$ of assuming H1 when H0 is true is no longer zero. According to the test, we should assume H1 whenever even a single photon is in mode $v_{10}$, meaning that to get $\alpha(N)$ we must sum Eq. (\ref{eq:binom}) over $0<k\leqslant N$, obtaining $\alpha(N)= 1 - (1-p_0)^N$. Similarly, $\beta(N)=\left(1 - p_x\right)^N$ is obtained by considering $k=0$ in Eq. (\ref{eq:binom}), corresponding to no photons in mode $v_{10}$. 
The total probability of error of the test in Eq.~\eqref{eq:original_test} is now
\begin{align}
\begin{split}
    P_{\textnormal{e}}(N) 
        = \frac{1}{2}\left(1 - (1-p_0)^N \right) + \frac{1}{2}\left(1 - p_x\right)^N.\label{eq:Pe_old}
\end{split}
\end{align}
Clearly, whenever $p_0\neq 0$, this approaches $1/2$ as $N\to\infty$: The introduction of any experimental error  changes the test from nearly optimal to as bad as flipping a coin.

\begin{figure}[!t]
    \centering
         \includegraphics[width=\textwidth/2]{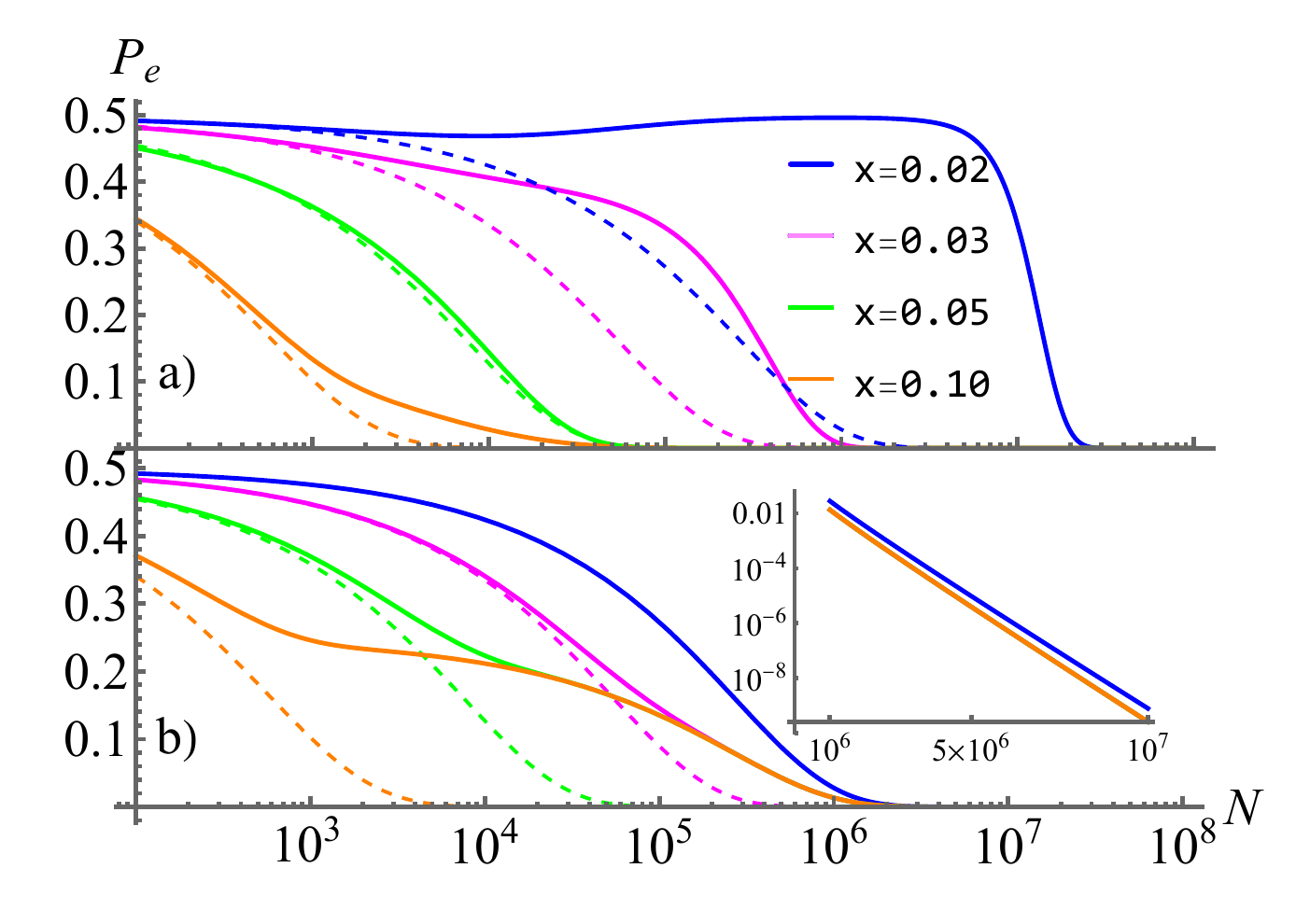}
        \caption{Probability of error versus $N$ assuming uniform crosstalk \cite{appendix} with $\epsilon^2=0.01$ and using Gaussian approximation to the binomial distribution. In both figures, from right to left, the dashed lines corresponds to the optimal test \eqref{eq:test_bi} for $x=0.02,\,0.03, \,0.05, \,0.10$ (blue, magenta, green, orange). a) Solid lines of corresponding colors stand for the distance-independent test (\ref{eq:tests_separation_independent}) for $\zeta=N^{4/5}/100$. 
        We see that the effectiveness of the distance-independent test is severely affected by the actual value of $x$, see \cite{appendix}. 
        b) Solid lines stand for the modified test (\ref{eq:algorithm_good}) with $x_{\min}=0.02$. Accordingly, the curve for $x=0.02$ (blue) provides an upper bound for all other curves. 
        The inset show how, in the $P_{\textnormal{e}}<0.05$ region, all solid curves in b) feature approximately the same rate of convergence.}
        \label{fig:P_N}  
\end{figure}

The failure of test (\ref{eq:original_test}) does not necessarily disqualify all separation-independent tests. 
In fact, it is possible to design a class of separation-independent tests for source discrimination with imperfect demultiplexers that yields $\lim_{N\to\infty}P_{\textnormal{e}}(N)=0$ \cite{appendix}. Such tests take the following form:
\begin{align} \label{eq:tests_separation_independent}
\begin{split}
    N_{10} \underset{H0}{\overset{H1}{\gtrless}} N p_0+\zeta(N),\\
\end{split}
\end{align}
where $\zeta(N)>0$ are $x$-independent functions increasing faster than $\sqrt{N}$, but slower than $N$. Probabilities of error for such tests can be analogously calculated using a binomial distribution as in case of Eq.~\eqref{eq:Pe_old} \cite{appendix}.
Note that the natural generalization of the test 
 (\ref{eq:original_test}) given by Eq.~\eqref{eq:tests_separation_independent} with $\zeta(N)=0$ results in the suboptimal $\lim_{N\to\infty}P_{\textnormal{e}}(N)=1/4$ \cite{appendix}. 
Unfortunately, the family (\ref{eq:tests_separation_independent}) appears only marginally more practical than the original test \eqref{eq:original_test}, as the corresponding rates of convergence of the probability of error to zero vary strongly with $x$, from nearly optimal to orders of magnitude worse (see Fig. \ref{fig:P_N} a)). 
Accordingly, despite the tests being $x$-independent, a sensible estimation of their probability of error, for given number $N$ of detected photons, would require some prior knowledge of $x$, severely undermining their practical utility.

These findings motivate us to search for more practical tests for source discrimination. To achieve this goal, we go back to the optimal likelihood-ratio test (\ref{eq:optimal_test}).  
Given that most information on the difference between the image of one source (H0) and that of two closely separated ones (H1) is contained in mode $v_{10}$, we rewrite Eq.~\eqref{eq:optimal_test}
 in terms of only two outcomes: photon in mode $v_{10}$ and photon in any other mode. This yields
\begin{align}
    \left(\frac{p(10|x)}{p(10|0)}\right)^{N_{10}}
        \left(\frac{1-p(10|x)}{1-p(10|0)}\right)^{N-N_{10}}  
        > 1,\label{eq:test_bi}
\end{align}
Solving for $N_{10}$ and expanding to second order in $x$ and $\epsilon$, we obtain
\begin{align} \label{eq:algorithm_good}
\begin{split}
    N_{10} \underset{H0}{\overset{H1}{\gtrless}} N\left(p_{0} + \gamma x^2/2\right),
\end{split}
\end{align}
where $\gamma = 1 - \mathcal{O}(\epsilon^2)$. 
The modified test (\ref{eq:algorithm_good}) is particularly appealing since (for small separations and weak crosstalk) it inherits the optimality of the likelihood ratio test while being simple and based on a single-mode measurement, like the tests Eq.~\eqref{eq:original_test} and \eqref{eq:tests_separation_independent} \cite{appendix}.
On the downside, Eq.~\eqref{eq:algorithm_good} is not separation-independent.

To obviate to this problem, let us replace $x$ on the right hand side of (\ref{eq:algorithm_good}) by some $x_{\min}$ and then test for the modified hypotheses: (H0) there is only one source or (H1) there are two sources with separation $x\geqslant x_{\min}$. 
In this case, we will still obtain $\lim_{N\to\infty}P_{\textnormal{e}}=0$ \cite{appendix}.  
Furthermore, for fixed $N$ and $x_{\min}$, $P_{\textnormal{e}}$ is decreasing with growing $x$, meaning that the probability of error of the algorithm is upper bounded by $P_{\textnormal{e}}$ calculated with $x=x_{\min}$. 
Therefore, the test \eqref{eq:algorithm_good} is semi-separation-independent, in the sense that for a fixed $x_{\min}$, despite being not optimal for every value of $x$, it allows for an easy access to a maximal probability of error independent to any a priori knowledge of the separation. This, in particular, avoids any possible underestimation of the actual error probability $P_e$. What is more, the considered test is near optimal for the modified hypothesis H1 among decision strategies which do not require estimation of the separation. All of this provides a reliable method of planning experiments: it is sufficient to set the minimal separation $x_{\min}$ into Eq.~\eqref{eq:algorithm_good} to determine the number of photons $N$ to detect to be sure not to exceed a pre-established maximal tolerable probability of error.
Fig. \ref{fig:P_N} b) shows a comparison between the real probability of error for this test and its upper bound.

\paragraph{Conclusions --}
We have scrutinized the effectiveness of realistic SPADE-based discrimination between one and two closely-separated light sources. 
We analytically showed that the presence of crosstalk heavily affects the probability of successful discrimination, causing it to scale suboptimally with the source separation $d$ even for relatively large values of $d$. 
Similarly, any crosstalk renders separation-independent hypothesis testing non-viable in practice. 
To remedy this, we proposed a simple semi-separation-independent algorithm based on the  likelihood-ratio test, which, even for imperfect demultiplexers, gives access to the maximal probability of error without requiring separation estimation. 

Our results suggest that it is mandatory to include the role of crosstalk, and other experimental imperfection, e.g. electronic noise \cite{Len,Giacomo}, in SPADE-based source discrimination.
In particular, it would be interesting to see how significant experimental imperfections are in multiple-hypotheses testing \cite{multihypothesis_testing_2022}, and when considering potentially unequal brightnesses of the sources \cite{Linowski}.

\paragraph{Acknowledgments --}
Project ApresSF is supported by National Science Centre in Poland (contract No.: UMO-2019/32/Z/ST2/00017) and ANR under the QuantERA programme, which has received funding from the European Union’s Horizon 2020 research and innovation programme.

\bibliography{PRL_manuscript}

\begin{thebibliography}{37}%
\makeatletter
\providecommand \@ifxundefined [1]{%
 \@ifx{#1\undefined}
}%
\providecommand \@ifnum [1]{%
 \ifnum #1\expandafter \@firstoftwo
 \else \expandafter \@secondoftwo
 \fi
}%
\providecommand \@ifx [1]{%
 \ifx #1\expandafter \@firstoftwo
 \else \expandafter \@secondoftwo
 \fi
}%
\providecommand \natexlab [1]{#1}%
\providecommand \enquote  [1]{``#1''}%
\providecommand \bibnamefont  [1]{#1}%
\providecommand \bibfnamefont [1]{#1}%
\providecommand \citenamefont [1]{#1}%
\providecommand \href@noop [0]{\@secondoftwo}%
\providecommand \href [0]{\begingroup \@sanitize@url \@href}%
\providecommand \@href[1]{\@@startlink{#1}\@@href}%
\providecommand \@@href[1]{\endgroup#1\@@endlink}%
\providecommand \@sanitize@url [0]{\catcode `\\12\catcode `\$12\catcode
  `\&12\catcode `\#12\catcode `\^12\catcode `\_12\catcode `\%12\relax}%
\providecommand \@@startlink[1]{}%
\providecommand \@@endlink[0]{}%
\providecommand \url  [0]{\begingroup\@sanitize@url \@url }%
\providecommand \@url [1]{\endgroup\@href {#1}{\urlprefix }}%
\providecommand \urlprefix  [0]{URL }%
\providecommand \Eprint [0]{\href }%
\providecommand \doibase [0]{http://dx.doi.org/}%
\providecommand \selectlanguage [0]{\@gobble}%
\providecommand \bibinfo  [0]{\@secondoftwo}%
\providecommand \bibfield  [0]{\@secondoftwo}%
\providecommand \translation [1]{[#1]}%
\providecommand \BibitemOpen [0]{}%
\providecommand \bibitemStop [0]{}%
\providecommand \bibitemNoStop [0]{.\EOS\space}%
\providecommand \EOS [0]{\spacefactor3000\relax}%
\providecommand \BibitemShut  [1]{\csname bibitem#1\endcsname}%
\let\auto@bib@innerbib\@empty
\bibitem [{\citenamefont {Harris}(1964)}]{Harris:64}%
  \BibitemOpen
  \bibfield  {author} {\bibinfo {author} {\bibfnamefont {J.~L.}\ \bibnamefont
  {Harris}},\ }\href {\doibase 10.1364/JOSA.54.000606} {\bibfield  {journal}
  {\bibinfo  {journal} {J. Opt. Soc. Am.}\ }\textbf {\bibinfo {volume} {54}},\
  \bibinfo {pages} {606} (\bibinfo {year} {1964})}\BibitemShut {NoStop}%
\bibitem [{\citenamefont {Acuna}\ and\ \citenamefont {Horowitz}(1997)}]{Exo_1}%
  \BibitemOpen
  \bibfield  {author} {\bibinfo {author} {\bibfnamefont {C.~O.}\ \bibnamefont
  {Acuna}}\ and\ \bibinfo {author} {\bibfnamefont {J.}~\bibnamefont
  {Horowitz}},\ }\href {\doibase 10.1080/02664769723620} {\bibfield  {journal}
  {\bibinfo  {journal} {J. App. Stat.}\ }\textbf {\bibinfo {volume} {24}},\
  \bibinfo {pages} {421} (\bibinfo {year} {1997})}\BibitemShut {NoStop}%
\bibitem [{\citenamefont {Shahram}\ and\ \citenamefont
  {Milanfar}(2006)}]{Exo_2}%
  \BibitemOpen
  \bibfield  {author} {\bibinfo {author} {\bibfnamefont {M.}~\bibnamefont
  {Shahram}}\ and\ \bibinfo {author} {\bibfnamefont {P.}~\bibnamefont
  {Milanfar}},\ }\href {\doibase 10.1109/TIT.2006.878180} {\bibfield  {journal}
  {\bibinfo  {journal} {IEEE Transactions on Information Theory}\ }\textbf
  {\bibinfo {volume} {52}},\ \bibinfo {pages} {3411} (\bibinfo {year}
  {2006})}\BibitemShut {NoStop}%
\bibitem [{\citenamefont {Labeyrie}\ \emph {et~al.}(2006)\citenamefont
  {Labeyrie}, \citenamefont {Lipson},\ and\ \citenamefont {Nisenson}}]{Exo_3}%
  \BibitemOpen
  \bibfield  {author} {\bibinfo {author} {\bibfnamefont {A.}~\bibnamefont
  {Labeyrie}}, \bibinfo {author} {\bibfnamefont {S.~G.}\ \bibnamefont
  {Lipson}}, \ and\ \bibinfo {author} {\bibfnamefont {P.}~\bibnamefont
  {Nisenson}},\ }\href {\doibase 10.1017/CBO9780511617638} {\emph {\bibinfo
  {title} {An Introduction to Optical Stellar Interferometry}}}\ (\bibinfo
  {publisher} {Cambridge University Press},\ \bibinfo {year}
  {2006})\BibitemShut {NoStop}%
\bibitem [{\citenamefont {{Wright}}\ and\ \citenamefont
  {{Gaudi}}(2013)}]{Exo_7}%
  \BibitemOpen
  \bibfield  {author} {\bibinfo {author} {\bibfnamefont {J.~T.}\ \bibnamefont
  {{Wright}}}\ and\ \bibinfo {author} {\bibfnamefont {B.~S.}\ \bibnamefont
  {{Gaudi}}},\ }in\ \href {\doibase 10.1007/978-94-007-5606-9_10} {\emph
  {\bibinfo {booktitle} {Planets, Stars and Stellar Systems. Volume 3: Solar
  and Stellar Planetary Systems}}},\ \bibinfo {editor} {edited by\ \bibinfo
  {editor} {\bibfnamefont {T.~D.}\ \bibnamefont {{Oswalt}}}, \bibinfo {editor}
  {\bibfnamefont {L.~M.}\ \bibnamefont {{French}}}, \ and\ \bibinfo {editor}
  {\bibfnamefont {P.}~\bibnamefont {{Kalas}}}}\ (\bibinfo  {publisher}
  {Springer, Dordrecht},\ \bibinfo {year} {2013})\ p.\ \bibinfo {pages}
  {489}\BibitemShut {NoStop}%
\bibitem [{\citenamefont {{Fischer}}\ \emph {et~al.}(2014)\citenamefont
  {{Fischer}}, \citenamefont {{Howard}}, \citenamefont {{Laughlin}},
  \citenamefont {{Macintosh}}, \citenamefont {{Mahadevan}}, \citenamefont
  {{Sahlmann}},\ and\ \citenamefont {{Yee}}}]{Exo_6}%
  \BibitemOpen
  \bibfield  {author} {\bibinfo {author} {\bibfnamefont {D.~A.}\ \bibnamefont
  {{Fischer}}}, \bibinfo {author} {\bibfnamefont {A.~W.}\ \bibnamefont
  {{Howard}}}, \bibinfo {author} {\bibfnamefont {G.~P.}\ \bibnamefont
  {{Laughlin}}}, \bibinfo {author} {\bibfnamefont {B.}~\bibnamefont
  {{Macintosh}}}, \bibinfo {author} {\bibfnamefont {S.}~\bibnamefont
  {{Mahadevan}}}, \bibinfo {author} {\bibfnamefont {J.}~\bibnamefont
  {{Sahlmann}}}, \ and\ \bibinfo {author} {\bibfnamefont {J.~C.}\ \bibnamefont
  {{Yee}}},\ }in\ \href {\doibase 10.2458/azu_uapress_9780816531240-ch031}
  {\emph {\bibinfo {booktitle} {Protostars and Planets VI}}},\ \bibinfo
  {editor} {edited by\ \bibinfo {editor} {\bibfnamefont {H.}~\bibnamefont
  {{Beuther}}}, \bibinfo {editor} {\bibfnamefont {R.~S.}\ \bibnamefont
  {{Klessen}}}, \bibinfo {editor} {\bibfnamefont {C.~P.}\ \bibnamefont
  {{Dullemond}}}, \ and\ \bibinfo {editor} {\bibfnamefont {T.}~\bibnamefont
  {{Henning}}}}\ (\bibinfo  {publisher} {University of Arizona Press},\
  \bibinfo {year} {2014})\ p.\ \bibinfo {pages} {715},\ \Eprint
  {http://arxiv.org/abs/1505.06869} {arXiv:1505.06869 [astro-ph.EP]}
  \BibitemShut {NoStop}%
\bibitem [{\citenamefont {Huang}\ and\ \citenamefont {Lupo}(2021)}]{Exo_4}%
  \BibitemOpen
  \bibfield  {author} {\bibinfo {author} {\bibfnamefont {Z.}~\bibnamefont
  {Huang}}\ and\ \bibinfo {author} {\bibfnamefont {C.}~\bibnamefont {Lupo}},\
  }\href {\doibase 10.1103/PhysRevLett.127.130502} {\bibfield  {journal}
  {\bibinfo  {journal} {Phys. Rev. Lett.}\ }\textbf {\bibinfo {volume} {127}},\
  \bibinfo {pages} {130502} (\bibinfo {year} {2021})}\BibitemShut {NoStop}%
\bibitem [{\citenamefont {Zanforlin}\ \emph {et~al.}(2022)\citenamefont
  {Zanforlin}, \citenamefont {Lupo}, \citenamefont {Connolly}, \citenamefont
  {Kok}, \citenamefont {Buller},\ and\ \citenamefont {Huang}}]{Exo_5}%
  \BibitemOpen
  \bibfield  {author} {\bibinfo {author} {\bibfnamefont {U.}~\bibnamefont
  {Zanforlin}}, \bibinfo {author} {\bibfnamefont {C.}~\bibnamefont {Lupo}},
  \bibinfo {author} {\bibfnamefont {P.~W.~R.}\ \bibnamefont {Connolly}},
  \bibinfo {author} {\bibfnamefont {P.}~\bibnamefont {Kok}}, \bibinfo {author}
  {\bibfnamefont {G.~S.}\ \bibnamefont {Buller}}, \ and\ \bibinfo {author}
  {\bibfnamefont {Z.}~\bibnamefont {Huang}},\ }\href {\doibase
  10.1038/s41467-022-32977-8} {\bibfield  {journal} {\bibinfo  {journal}
  {Nature Communications}\ }\textbf {\bibinfo {volume} {13}},\ \bibinfo {pages}
  {5373} (\bibinfo {year} {2022})}\BibitemShut {NoStop}%
\bibitem [{\citenamefont {Lee}\ \emph {et~al.}(2012)\citenamefont {Lee},
  \citenamefont {Shin}, \citenamefont {Lee},\ and\ \citenamefont
  {Bustamante}}]{Micro_2}%
  \BibitemOpen
  \bibfield  {author} {\bibinfo {author} {\bibfnamefont {S.-H.}\ \bibnamefont
  {Lee}}, \bibinfo {author} {\bibfnamefont {J.~Y.}\ \bibnamefont {Shin}},
  \bibinfo {author} {\bibfnamefont {A.}~\bibnamefont {Lee}}, \ and\ \bibinfo
  {author} {\bibfnamefont {C.}~\bibnamefont {Bustamante}},\ }\href@noop {}
  {\bibfield  {journal} {\bibinfo  {journal} {Proceedings of the National
  Academy of Sciences}\ }\textbf {\bibinfo {volume} {109}},\ \bibinfo {pages}
  {17436} (\bibinfo {year} {2012})}\BibitemShut {NoStop}%
\bibitem [{\citenamefont {Nan}\ \emph {et~al.}(2013)\citenamefont {Nan},
  \citenamefont {Collisson}, \citenamefont {Lewis}, \citenamefont {Huang},
  \citenamefont {Tamg{\"u}ney}, \citenamefont {Liphardt}, \citenamefont
  {McCormick}, \citenamefont {Gray},\ and\ \citenamefont {Chu}}]{Micro_1}%
  \BibitemOpen
  \bibfield  {author} {\bibinfo {author} {\bibfnamefont {X.}~\bibnamefont
  {Nan}}, \bibinfo {author} {\bibfnamefont {E.~A.}\ \bibnamefont {Collisson}},
  \bibinfo {author} {\bibfnamefont {S.}~\bibnamefont {Lewis}}, \bibinfo
  {author} {\bibfnamefont {J.}~\bibnamefont {Huang}}, \bibinfo {author}
  {\bibfnamefont {T.~M.}\ \bibnamefont {Tamg{\"u}ney}}, \bibinfo {author}
  {\bibfnamefont {J.~T.}\ \bibnamefont {Liphardt}}, \bibinfo {author}
  {\bibfnamefont {F.}~\bibnamefont {McCormick}}, \bibinfo {author}
  {\bibfnamefont {J.~W.}\ \bibnamefont {Gray}}, \ and\ \bibinfo {author}
  {\bibfnamefont {S.}~\bibnamefont {Chu}},\ }\href@noop {} {\bibfield
  {journal} {\bibinfo  {journal} {Proceedings of the National Academy of
  Sciences}\ }\textbf {\bibinfo {volume} {110}},\ \bibinfo {pages} {18519}
  (\bibinfo {year} {2013})}\BibitemShut {NoStop}%
\bibitem [{\citenamefont {Baddeley}\ and\ \citenamefont
  {Bewersdorf}(2018)}]{Micro_3}%
  \BibitemOpen
  \bibfield  {author} {\bibinfo {author} {\bibfnamefont {D.}~\bibnamefont
  {Baddeley}}\ and\ \bibinfo {author} {\bibfnamefont {J.}~\bibnamefont
  {Bewersdorf}},\ }\href {\doibase 10.1146/annurev-biochem-060815-014801}
  {\bibfield  {journal} {\bibinfo  {journal} {Annual Review of Biochemistry}\
  }\textbf {\bibinfo {volume} {87}},\ \bibinfo {pages} {965} (\bibinfo {year}
  {2018})}\BibitemShut {NoStop}%
\bibitem [{\citenamefont {den Dekker}\ and\ \citenamefont {van~den
  Bos}(1997)}]{resolution_survey_denDekker_1997}%
  \BibitemOpen
  \bibfield  {author} {\bibinfo {author} {\bibfnamefont {A.~J.}\ \bibnamefont
  {den Dekker}}\ and\ \bibinfo {author} {\bibfnamefont {A.}~\bibnamefont
  {van~den Bos}},\ }\href {\doibase 10.1364/JOSAA.14.000547} {\bibfield
  {journal} {\bibinfo  {journal} {J. Opt. Soc. Am. A}\ }\textbf {\bibinfo
  {volume} {14}},\ \bibinfo {pages} {547} (\bibinfo {year} {1997})}\BibitemShut
  {NoStop}%
\bibitem [{\citenamefont {Goodman}(2005)}]{Fourier_optics_Goodman_2005}%
  \BibitemOpen
  \bibfield  {author} {\bibinfo {author} {\bibfnamefont {J.~W.}\ \bibnamefont
  {Goodman}},\ }\href@noop {} {\emph {\bibinfo {title} {Introduction to Fourier
  optics}}}\ (\bibinfo  {publisher} {Roberts and Company Publishers},\ \bibinfo
  {year} {2005})\BibitemShut {NoStop}%
\bibitem [{\citenamefont {Helstrom}(1973)}]{Helstrom}%
  \BibitemOpen
  \bibfield  {author} {\bibinfo {author} {\bibfnamefont {C.}~\bibnamefont
  {Helstrom}},\ }\href {\doibase 10.1109/TIT.1973.1055052} {\bibfield
  {journal} {\bibinfo  {journal} {IEEE Transactions on Information Theory}\
  }\textbf {\bibinfo {volume} {19}},\ \bibinfo {pages} {389} (\bibinfo {year}
  {1973})}\BibitemShut {NoStop}%
\bibitem [{\citenamefont {Tsang}\ \emph {et~al.}(2016)\citenamefont {Tsang},
  \citenamefont {Nair},\ and\ \citenamefont {Lu}}]{superresolution_Tsang_2016}%
  \BibitemOpen
  \bibfield  {author} {\bibinfo {author} {\bibfnamefont {M.}~\bibnamefont
  {Tsang}}, \bibinfo {author} {\bibfnamefont {R.}~\bibnamefont {Nair}}, \ and\
  \bibinfo {author} {\bibfnamefont {X.-M.}\ \bibnamefont {Lu}},\ }\href
  {\doibase 10.1103/PhysRevX.6.031033} {\bibfield  {journal} {\bibinfo
  {journal} {Phys. Rev. X}\ }\textbf {\bibinfo {volume} {6}},\ \bibinfo {pages}
  {031033} (\bibinfo {year} {2016})}\BibitemShut {NoStop}%
\bibitem [{\citenamefont {Tsang}(2019)}]{superresolution_starlight_Tsang_2019}%
  \BibitemOpen
  \bibfield  {author} {\bibinfo {author} {\bibfnamefont {M.}~\bibnamefont
  {Tsang}},\ }\href {\doibase 10.1080/00107514.2020.1736375} {\bibfield
  {journal} {\bibinfo  {journal} {Contemp. Phys.}\ }\textbf {\bibinfo {volume}
  {60}},\ \bibinfo {pages} {279} (\bibinfo {year} {2019})}\BibitemShut
  {NoStop}%
\bibitem [{\citenamefont {Lu}\ \emph {et~al.}(2018)\citenamefont {Lu},
  \citenamefont {Krovi}, \citenamefont {Nair}, \citenamefont {Guha},\ and\
  \citenamefont {Shapiro}}]{HYP_1}%
  \BibitemOpen
  \bibfield  {author} {\bibinfo {author} {\bibfnamefont {X.-M.}\ \bibnamefont
  {Lu}}, \bibinfo {author} {\bibfnamefont {H.}~\bibnamefont {Krovi}}, \bibinfo
  {author} {\bibfnamefont {R.}~\bibnamefont {Nair}}, \bibinfo {author}
  {\bibfnamefont {S.}~\bibnamefont {Guha}}, \ and\ \bibinfo {author}
  {\bibfnamefont {J.~H.}\ \bibnamefont {Shapiro}},\ }\href {\doibase
  10.1038/s41534-018-0114-y} {\bibfield  {journal} {\bibinfo  {journal} {npj
  Quantum Information}\ }\textbf {\bibinfo {volume} {4}},\ \bibinfo {pages}
  {64} (\bibinfo {year} {2018})}\BibitemShut {NoStop}%
\bibitem [{\citenamefont {Grace}\ and\ \citenamefont
  {Guha}(2022{\natexlab{a}})}]{no_point}%
  \BibitemOpen
  \bibfield  {author} {\bibinfo {author} {\bibfnamefont {M.~R.}\ \bibnamefont
  {Grace}}\ and\ \bibinfo {author} {\bibfnamefont {S.}~\bibnamefont {Guha}},\
  }\href {\doibase 10.1103/PhysRevLett.129.180502} {\bibfield  {journal}
  {\bibinfo  {journal} {Phys. Rev. Lett.}\ }\textbf {\bibinfo {volume} {129}},\
  \bibinfo {pages} {180502} (\bibinfo {year} {2022}{\natexlab{a}})}\BibitemShut
  {NoStop}%
\bibitem [{\citenamefont {Giovannetti}\ \emph {et~al.}(2011)\citenamefont
  {Giovannetti}, \citenamefont {Lloyd},\ and\ \citenamefont
  {Maccone}}]{noon_advances}%
  \BibitemOpen
  \bibfield  {author} {\bibinfo {author} {\bibfnamefont {V.}~\bibnamefont
  {Giovannetti}}, \bibinfo {author} {\bibfnamefont {S.}~\bibnamefont {Lloyd}},
  \ and\ \bibinfo {author} {\bibfnamefont {L.}~\bibnamefont {Maccone}},\ }\href
  {\doibase 10.1038/nphoton.2011.35} {\bibfield  {journal} {\bibinfo  {journal}
  {Nat. Photonics}\ }\textbf {\bibinfo {volume} {5}},\ \bibinfo {pages} {2733}
  (\bibinfo {year} {2011})}\BibitemShut {NoStop}%
\bibitem [{\citenamefont {Ko{\l}ody{\'{n}}ski}\ and\ \citenamefont
  {Demkowicz-Dobrza{\'{n}}ski}(2013)}]{metrology_noise_Kolodynski_2013}%
  \BibitemOpen
  \bibfield  {author} {\bibinfo {author} {\bibfnamefont {J.}~\bibnamefont
  {Ko{\l}ody{\'{n}}ski}}\ and\ \bibinfo {author} {\bibfnamefont
  {R.}~\bibnamefont {Demkowicz-Dobrza{\'{n}}ski}},\ }\href {\doibase
  10.1088/1367-2630/15/7/073043} {\bibfield  {journal} {\bibinfo  {journal}
  {New J. Phys.}\ }\textbf {\bibinfo {volume} {15}},\ \bibinfo {pages} {073043}
  (\bibinfo {year} {2013})}\BibitemShut {NoStop}%
\bibitem [{\citenamefont {Nichols}\ \emph {et~al.}(2016)\citenamefont
  {Nichols}, \citenamefont {Bromley}, \citenamefont {Correa},\ and\
  \citenamefont {Adesso}}]{metrology_noise_Nichols_2016}%
  \BibitemOpen
  \bibfield  {author} {\bibinfo {author} {\bibfnamefont {R.}~\bibnamefont
  {Nichols}}, \bibinfo {author} {\bibfnamefont {T.~R.}\ \bibnamefont
  {Bromley}}, \bibinfo {author} {\bibfnamefont {L.~A.}\ \bibnamefont {Correa}},
  \ and\ \bibinfo {author} {\bibfnamefont {G.}~\bibnamefont {Adesso}},\ }\href
  {\doibase 10.1103/PhysRevA.94.042101} {\bibfield  {journal} {\bibinfo
  {journal} {Phys. Rev. A}\ }\textbf {\bibinfo {volume} {94}},\ \bibinfo
  {pages} {042101} (\bibinfo {year} {2016})}\BibitemShut {NoStop}%
\bibitem [{\citenamefont {Oh}\ \emph {et~al.}(2021)\citenamefont {Oh},
  \citenamefont {Zhou}, \citenamefont {Wong},\ and\ \citenamefont
  {Jiang}}]{noise_Oh_2021}%
  \BibitemOpen
  \bibfield  {author} {\bibinfo {author} {\bibfnamefont {C.}~\bibnamefont
  {Oh}}, \bibinfo {author} {\bibfnamefont {S.}~\bibnamefont {Zhou}}, \bibinfo
  {author} {\bibfnamefont {Y.}~\bibnamefont {Wong}}, \ and\ \bibinfo {author}
  {\bibfnamefont {L.}~\bibnamefont {Jiang}},\ }\href {\doibase
  10.1103/PhysRevLett.126.120502} {\bibfield  {journal} {\bibinfo  {journal}
  {Phys. Rev. Lett.}\ }\textbf {\bibinfo {volume} {126}},\ \bibinfo {pages}
  {120502} (\bibinfo {year} {2021})}\BibitemShut {NoStop}%
\bibitem [{\citenamefont {Sorelli}\ \emph
  {et~al.}(2021{\natexlab{a}})\citenamefont {Sorelli}, \citenamefont {Gessner},
  \citenamefont {Walschaers},\ and\ \citenamefont
  {Treps}}]{Sorelli_practical_superresolution_2021}%
  \BibitemOpen
  \bibfield  {author} {\bibinfo {author} {\bibfnamefont {G.}~\bibnamefont
  {Sorelli}}, \bibinfo {author} {\bibfnamefont {M.}~\bibnamefont {Gessner}},
  \bibinfo {author} {\bibfnamefont {M.}~\bibnamefont {Walschaers}}, \ and\
  \bibinfo {author} {\bibfnamefont {N.}~\bibnamefont {Treps}},\ }\href
  {\doibase 10.1103/PhysRevLett.127.123604} {\bibfield  {journal} {\bibinfo
  {journal} {Phys. Rev. Lett.}\ }\textbf {\bibinfo {volume} {127}},\ \bibinfo
  {pages} {123604} (\bibinfo {year} {2021}{\natexlab{a}})}\BibitemShut
  {NoStop}%
\bibitem [{\citenamefont {Lupo}(2020)}]{imaging_noisy_Lupo_2020}%
  \BibitemOpen
  \bibfield  {author} {\bibinfo {author} {\bibfnamefont {C.}~\bibnamefont
  {Lupo}},\ }\href {\doibase 10.1103/PhysRevA.101.022323} {\bibfield  {journal}
  {\bibinfo  {journal} {Phys. Rev. A}\ }\textbf {\bibinfo {volume} {101}},\
  \bibinfo {pages} {022323} (\bibinfo {year} {2020})}\BibitemShut {NoStop}%
\bibitem [{\citenamefont {Len}\ \emph {et~al.}(2020{\natexlab{a}})\citenamefont
  {Len}, \citenamefont {Datta}, \citenamefont {Parniak},\ and\ \citenamefont
  {Banaszek}}]{superresolution_limits_SPADE_Len_2020}%
  \BibitemOpen
  \bibfield  {author} {\bibinfo {author} {\bibfnamefont {Y.~L.}\ \bibnamefont
  {Len}}, \bibinfo {author} {\bibfnamefont {C.}~\bibnamefont {Datta}}, \bibinfo
  {author} {\bibfnamefont {M.}~\bibnamefont {Parniak}}, \ and\ \bibinfo
  {author} {\bibfnamefont {K.}~\bibnamefont {Banaszek}},\ }\href {\doibase
  10.1142/S0219749919410156} {\bibfield  {journal} {\bibinfo  {journal} {Int.
  J. Quantum Inf.}\ }\textbf {\bibinfo {volume} {18}},\ \bibinfo {pages}
  {1941015} (\bibinfo {year} {2020}{\natexlab{a}})}\BibitemShut {NoStop}%
\bibitem [{\citenamefont {de~Almeida}\ \emph {et~al.}(2021)\citenamefont
  {de~Almeida}, \citenamefont {Ko\l{}ody\ifmmode~\acute{n}\else \'{n}\fi{}ski},
  \citenamefont {Hirche}, \citenamefont {Lewenstein},\ and\ \citenamefont
  {Skotiniotis}}]{resolution_misalingment_Almeida_2021}%
  \BibitemOpen
  \bibfield  {author} {\bibinfo {author} {\bibfnamefont {J.~O.}\ \bibnamefont
  {de~Almeida}}, \bibinfo {author} {\bibfnamefont {J.}~\bibnamefont
  {Ko\l{}ody\ifmmode~\acute{n}\else \'{n}\fi{}ski}}, \bibinfo {author}
  {\bibfnamefont {C.}~\bibnamefont {Hirche}}, \bibinfo {author} {\bibfnamefont
  {M.}~\bibnamefont {Lewenstein}}, \ and\ \bibinfo {author} {\bibfnamefont
  {M.}~\bibnamefont {Skotiniotis}},\ }\href {\doibase
  10.1103/PhysRevA.103.022406} {\bibfield  {journal} {\bibinfo  {journal}
  {Phys. Rev. A}\ }\textbf {\bibinfo {volume} {103}},\ \bibinfo {pages}
  {022406} (\bibinfo {year} {2021})}\BibitemShut {NoStop}%
\bibitem [{\citenamefont {Grace}\ \emph {et~al.}(2020)\citenamefont {Grace},
  \citenamefont {Dutton}, \citenamefont {Ashok},\ and\ \citenamefont
  {Guha}}]{superresolution_no_location_Grace_2020}%
  \BibitemOpen
  \bibfield  {author} {\bibinfo {author} {\bibfnamefont {M.~R.}\ \bibnamefont
  {Grace}}, \bibinfo {author} {\bibfnamefont {Z.}~\bibnamefont {Dutton}},
  \bibinfo {author} {\bibfnamefont {A.}~\bibnamefont {Ashok}}, \ and\ \bibinfo
  {author} {\bibfnamefont {S.}~\bibnamefont {Guha}},\ }\href {\doibase
  10.1364/JOSAA.392116} {\bibfield  {journal} {\bibinfo  {journal} {J. Opt.
  Soc. Am. A}\ }\textbf {\bibinfo {volume} {37}},\ \bibinfo {pages} {1288}
  (\bibinfo {year} {2020})}\BibitemShut {NoStop}%
\bibitem [{\citenamefont {Gessner}\ \emph {et~al.}(2020)\citenamefont
  {Gessner}, \citenamefont {Fabre},\ and\ \citenamefont {Treps}}]{Manuel}%
  \BibitemOpen
  \bibfield  {author} {\bibinfo {author} {\bibfnamefont {M.}~\bibnamefont
  {Gessner}}, \bibinfo {author} {\bibfnamefont {C.}~\bibnamefont {Fabre}}, \
  and\ \bibinfo {author} {\bibfnamefont {N.}~\bibnamefont {Treps}},\ }\href
  {\doibase 10.1103/PhysRevLett.125.100501} {\bibfield  {journal} {\bibinfo
  {journal} {Phys. Rev. Lett.}\ }\textbf {\bibinfo {volume} {125}},\ \bibinfo
  {pages} {100501} (\bibinfo {year} {2020})}\BibitemShut {NoStop}%
\bibitem [{\citenamefont {Sorelli}\ \emph
  {et~al.}(2021{\natexlab{b}})\citenamefont {Sorelli}, \citenamefont {Gessner},
  \citenamefont {Walschaers},\ and\ \citenamefont {Treps}}]{Giacomo}%
  \BibitemOpen
  \bibfield  {author} {\bibinfo {author} {\bibfnamefont {G.}~\bibnamefont
  {Sorelli}}, \bibinfo {author} {\bibfnamefont {M.}~\bibnamefont {Gessner}},
  \bibinfo {author} {\bibfnamefont {M.}~\bibnamefont {Walschaers}}, \ and\
  \bibinfo {author} {\bibfnamefont {N.}~\bibnamefont {Treps}},\ }\href
  {\doibase 10.1103/PhysRevA.104.033515} {\bibfield  {journal} {\bibinfo
  {journal} {Phys. Rev. A}\ }\textbf {\bibinfo {volume} {104}},\ \bibinfo
  {pages} {033515} (\bibinfo {year} {2021}{\natexlab{b}})}\BibitemShut
  {NoStop}%
\bibitem [{\citenamefont {Linowski}\ \emph {et~al.}(2022)\citenamefont
  {Linowski}, \citenamefont {Schlichtholz}, \citenamefont {Sorelli},
  \citenamefont {Gessner}, \citenamefont {Walschaers}, \citenamefont {Treps},\
  and\ \citenamefont {Rudnicki}}]{Linowski}%
  \BibitemOpen
  \bibfield  {author} {\bibinfo {author} {\bibfnamefont {T.}~\bibnamefont
  {Linowski}}, \bibinfo {author} {\bibfnamefont {K.}~\bibnamefont
  {Schlichtholz}}, \bibinfo {author} {\bibfnamefont {G.}~\bibnamefont
  {Sorelli}}, \bibinfo {author} {\bibfnamefont {M.}~\bibnamefont {Gessner}},
  \bibinfo {author} {\bibfnamefont {M.}~\bibnamefont {Walschaers}}, \bibinfo
  {author} {\bibfnamefont {N.}~\bibnamefont {Treps}}, \ and\ \bibinfo {author}
  {\bibfnamefont {{\L}.}~\bibnamefont {Rudnicki}},\ }\href {\doibase
  10.48550/ARXIV.2211.09157} {\enquote {\bibinfo {title} {Application range of
  crosstalk-affected spatial demultiplexing for resolving separations between
  unbalanced sources},}\ } (\bibinfo {year} {2022})\BibitemShut {NoStop}%
\bibitem [{\citenamefont {Goodman}(1985)}]{goodman1985}%
  \BibitemOpen
  \bibfield  {author} {\bibinfo {author} {\bibfnamefont {J.~W.}\ \bibnamefont
  {Goodman}},\ }\href@noop {} {\emph {\bibinfo {title} {Statistical optics}}}\
  (\bibinfo  {publisher} {Wiley},\ \bibinfo {address} {New York},\ \bibinfo
  {year} {1985})\BibitemShut {NoStop}%
\bibitem [{\citenamefont {Chernoff}(1952)}]{Chernoff}%
  \BibitemOpen
  \bibfield  {author} {\bibinfo {author} {\bibfnamefont {H.}~\bibnamefont
  {Chernoff}},\ }\href {http://dml.mathdoc.fr/item/1177729330} {\bibfield
  {journal} {\bibinfo  {journal} {Ann. Math. Statist.}\ }\textbf {\bibinfo
  {volume} {23}},\ \bibinfo {pages} {493} (\bibinfo {year} {1952})}\BibitemShut
  {NoStop}%
\bibitem [{\citenamefont {Ogawa}\ and\ \citenamefont
  {Hayashi}(2004)}]{Q_Chernoff}%
  \BibitemOpen
  \bibfield  {author} {\bibinfo {author} {\bibfnamefont {T.}~\bibnamefont
  {Ogawa}}\ and\ \bibinfo {author} {\bibfnamefont {M.}~\bibnamefont
  {Hayashi}},\ }\href {\doibase 10.1109/TIT.2004.828155} {\bibfield  {journal}
  {\bibinfo  {journal} {IEEE Transactions on Information Theory}\ }\textbf
  {\bibinfo {volume} {50}},\ \bibinfo {pages} {1368} (\bibinfo {year}
  {2004})}\BibitemShut {NoStop}%
\bibitem [{app()}]{appendix}%
  \BibitemOpen
  \href@noop {} {\ }\bibinfo {note} {For more details on crosstalk, detailed
  proofs of our claims and additional figures, see Supplementary
  Material.}\BibitemShut {Stop}%
\bibitem [{\citenamefont {Van~Trees}(2001)}]{Van_Trees}%
  \BibitemOpen
  \bibfield  {author} {\bibinfo {author} {\bibfnamefont {H.~L.}\ \bibnamefont
  {Van~Trees}},\ }\href@noop {} {\emph {\bibinfo {title} {Detection,
  Estimation, and Modulation Theory, Part I.}}}\ (\bibinfo  {publisher} {John
  Wiley and Sons, Inc.},\ \bibinfo {year} {2001})\BibitemShut {NoStop}%
\bibitem [{\citenamefont {Len}\ \emph {et~al.}(2020{\natexlab{b}})\citenamefont
  {Len}, \citenamefont {Datta}, \citenamefont {Parniak},\ and\ \citenamefont
  {Banaszek}}]{Len}%
  \BibitemOpen
  \bibfield  {author} {\bibinfo {author} {\bibfnamefont {Y.~L.}\ \bibnamefont
  {Len}}, \bibinfo {author} {\bibfnamefont {C.}~\bibnamefont {Datta}}, \bibinfo
  {author} {\bibfnamefont {M.}~\bibnamefont {Parniak}}, \ and\ \bibinfo
  {author} {\bibfnamefont {K.}~\bibnamefont {Banaszek}},\ }\href {\doibase
  10.1142/S0219749919410156} {\bibfield  {journal} {\bibinfo  {journal}
  {International Journal of Quantum Information}\ }\textbf {\bibinfo {volume}
  {18}},\ \bibinfo {pages} {1941015} (\bibinfo {year}
  {2020}{\natexlab{b}})}\BibitemShut {NoStop}%
\bibitem [{\citenamefont {Grace}\ and\ \citenamefont
  {Guha}(2022{\natexlab{b}})}]{multihypothesis_testing_2022}%
  \BibitemOpen
  \bibfield  {author} {\bibinfo {author} {\bibfnamefont {M.~R.}\ \bibnamefont
  {Grace}}\ and\ \bibinfo {author} {\bibfnamefont {S.}~\bibnamefont {Guha}},\
  }\href {\doibase 10.1103/PhysRevLett.129.180502} {\bibfield  {journal}
  {\bibinfo  {journal} {Phys. Rev. Lett.}\ }\textbf {\bibinfo {volume} {129}},\
  \bibinfo {pages} {180502} (\bibinfo {year} {2022}{\natexlab{b}})}\BibitemShut
  {NoStop}%
\end{thebibliography}%

\clearpage

\begin{center}\Large\textbf{Supplemental Material}\end{center}\normalsize
In Section I, we recall the basic information about the crosstalk-affected SPADE measurement. In Section II, we calculate the Chernoff exponent for small separations for crosstalk-affected SPADE and ideal direct imaging. In Section III, we derive and discuss the separation-independent tests from the main text. Finally, in Section IV, we derive our semi-separation-independent test and provide its intuitive explanation.

\setcounter{section}{0}
\section{I. Crosstalk-affected SPADE}\label{app:SPADE}
\setcounter{section}{0}
\setcounter{equation}{0}
\renewcommand{\theequation}{S\arabic{equation}}
\setcounter{figure}{0}
\renewcommand{\thefigure}{S\arabic{figure}}
Let us recall some basic information about the SPADE-based measurement in the presence of crosstalk \cite{Manuel}. The probability of measuring a photon in the $k$-th mode when performing ideal photon counting in some basis $v_{k}$ upon the state from Eq.~(1) in the main text is given by:
\begin{align} \label{eq:p_SPADE_v}
    p(nm|d)  = 
        \frac{1}{2}\left(|f_{+ k}(d)|^2 + |f_{- k}(d)|^2\right),
\end{align}
where 
\begin{align} \label{eq:f_nm,kl}
\begin{split}
    f_{\pm k}(d) &= \int_{\mathbb{R}^2} d^2\vec{r} \, v_{k}^*(\vec{r}) \, u_{ 00}(\vec{r}\mp\vec{d})
\end{split}
\end{align}
are the overlap integrals of measurement basis function $v_{nm}(\vec r)$ and the spatial field distribution resulting from the source positioned at $\pm\vec{r}_0$. 

Consider the basis given by the Hermite-Gauss modes:
\begin{align} \label{eq:HG_modes}
    u_{nm}(\vec{r}) \coloneqq 
        \frac{H_n(\sqrt{2}r_x/w)H_m(\sqrt{2}r_y/w)}{w\sqrt{2^{n+m-1}\pi n! m!}}
        e^{-(r_x^2+r_y^2)/w^2},
\end{align}
where $\vec{r}=(r_x,r_y)$ and $H_n(z)\coloneqq (-1)^n e^{z^2} \partial_z^n e^{-z^2}$ are the Hermite polynomials and $w$ is the width of the point spread function of the imaging system. 
For ideal measurements in this basis, i.e. $v_k(\vec r)=u_{nm}(\vec r)$, the overlap integrals \eqref{eq:f_nm,kl} equal \cite{Manuel}
\begin{multline} \label{eq:beta_nm}
    f_{\pm nm}(d)=\beta_{\pm nm}(d) = \\=\frac{1}{\sqrt{n!}}\left(\pm \frac{d}{2w}\right)^{n} 
        \, e^{-d^2/8w^2}\delta_{0m},
\end{multline}
where $\delta_{0m}$ stands for Kronecker delta. 

However, as discussed in the main text, in reality, SPADE is subject to crosstalk. This changes the overlap integrals \eqref{eq:f_nm,kl} to
\begin{align} \label{eq:gamma_nm} 
    f_{\pm nm}(d) = \sum_{k,l=0}^{D-1} C_{nm,kl} \beta_{\pm kl},
\end{align}
where $C$ is the crosstalk matrix. Note that, due to the restriction to finite $D$, corresponding measurement probabilities (\ref{eq:p_SPADE_v}) have to be renormalized:
\begin{equation} 
    p(nm|d,D)=\frac{ p(nm|d)}{ \sum_{n,m}^{D-1}p(nm|d)}.
\end{equation}


In a well-designed experiment, crosstalk is relatively weak. Because $C$ is unitary, it can be written as \cite{Manuel}
\begin{align} \label{eq:crosstalk_matrix_unitary_representation}
    C = e^{-i\mu \vec\lambda\cdot\vec{G}},
\end{align}
where $\mu\geqslant 0$, $\vec{G}$ denotes a vector of all $D^4-1$ generalized Gell-Mann matrices of size $D^2\times D^2$ and  $\vec{\lambda}\in\mathbb{R}^{D^4-1}$ is some normalized vector. If $\mu\ll1$ then matrix $C$ is close to identity, thus describing small imperfections of the measurement apparatus.

The severity of imperfections can be quantified by the \emph{crosstalk strength} \cite{Manuel}:
\begin{align} \label{eq:crosstalk_strength}
    \epsilon^2\coloneqq \frac{1}{D^2(D^2-1)}\sum_{\substack{n,m,k,l=0\\nm\neq kl}}^{D-1} |C_{nm,kl}|^2.
\end{align}
For weak crosstalk matrices, that is, for $\mu\ll1$, one can find that on average \cite{Linowski}:
\begin{align} \label{eq:p_c(mu)_avg}
    \epsilon^2(\mu)\approx \frac{2}{D^4-1}\mu^2.
\end{align}
Due to its intuitive interpretation as the average probability of crosstalk per mode and its accessibility in experiment, we use the crosstalk strength $\epsilon^2$ rather than the abstract parameter $\mu$. This provides a method for generating random generic crosstalk matrices, by randomly choosing $\lambda$ from the uniform distribution on the $(D^4-1)$-sphere.

For qualitative considerations, a simplified     \emph{uniform crosstalk} model can be used to show the general behaviour of the SPADE-based measurement:
\begin{multline}
\label{eq:uniform_crosstalk_single_eq}
    (C_{\textnormal{u}})_{nm,kl} =\\ \delta_{nk}\delta_{ml}\sqrt{1-(D^2-1)\epsilon^2} + \left(1-\delta_{nk}\delta_{ml}\right) \epsilon.
\end{multline}
Note that the uniform crosstalk matrix \eqref{eq:uniform_crosstalk_single_eq} is fully defined by the number of measured modes $D^2-1$ and the crosstalk strength $\epsilon^2$.

\section{II. Chernoff exponent for small separations}
Here, we calculate the approximate Chernoff exponent for the crosstalk-affected SPADE and direct imaging in the limit of small separations.

\subsection{A. SPADE}
For SPADE, we begin by observing that
\begin{align}
    \xi&=-\ln(\min_{0\leq s\leq1}\sum_{n,m=0}^{D-1}p(nm|0,D)^sp(nm|d,D)^{1-s}) \nonumber\\
        &\equiv-\ln(\min_{0\leq s\leq1}Q_s).\label{eq:chernoff2}
\end{align}
Because, in our case of interest, both $x\ll1$ and $\epsilon\ll1$, we can expand $Q_s$ in these parameters. 

Observe that, as with any function of two variables, the result may in principle depend on the order of expansion. This is indeed what we find here:
\begin{align} \label{eq:F_limits}
    Q_s\approx
    \begin{cases}
        1+(s-1+(p_0/x^2)^s)x^2, & x\gg \epsilon, \\   
         1+(s-1) s x^4/(2 p_0), & x\ll \epsilon.
    \end{cases}
\end{align}
Here, the upper line was obtained by expanding in $\epsilon$ first, which corresponds to $x\gg \epsilon$, while the bottom line was obtained by expanding in $x$ first, which corresponds to $x\ll \epsilon$. 
Taking the derivative with respect to $s$ of Eq.~\eqref{eq:F_limits}, equating it to zero and solving for $s$, we obtain that the minimum of $Q_s$ is at
\begin{align} \label{eq:s_min}
    s_{\min}=
    \begin{cases}
        -\ln{(-\ln{(x^2/\epsilon^2)})}/\ln{(x^2/\epsilon^2)}, & x\gg \epsilon, \\   
        1/2, & x\ll \epsilon.
    \end{cases}
\end{align} 
Substituting this into the approximate Chernoff exponent (\ref{eq:chernoff2}) and once again performing series expansion in $x$ yields the formula (5) from the main text:
\begin{align} 
    \xi\approx
    \begin{cases}
        \left\{1 - \left[\ln\ln q(x)-1\right]/\ln q(x) \right\} x^2 & x\gg \epsilon, \\   
        x^4/(8 p_0), & x\ll \epsilon.
    \end{cases}
\end{align}

\begin{figure}[!t]
\centering
         \includegraphics[width=0.49\textwidth]{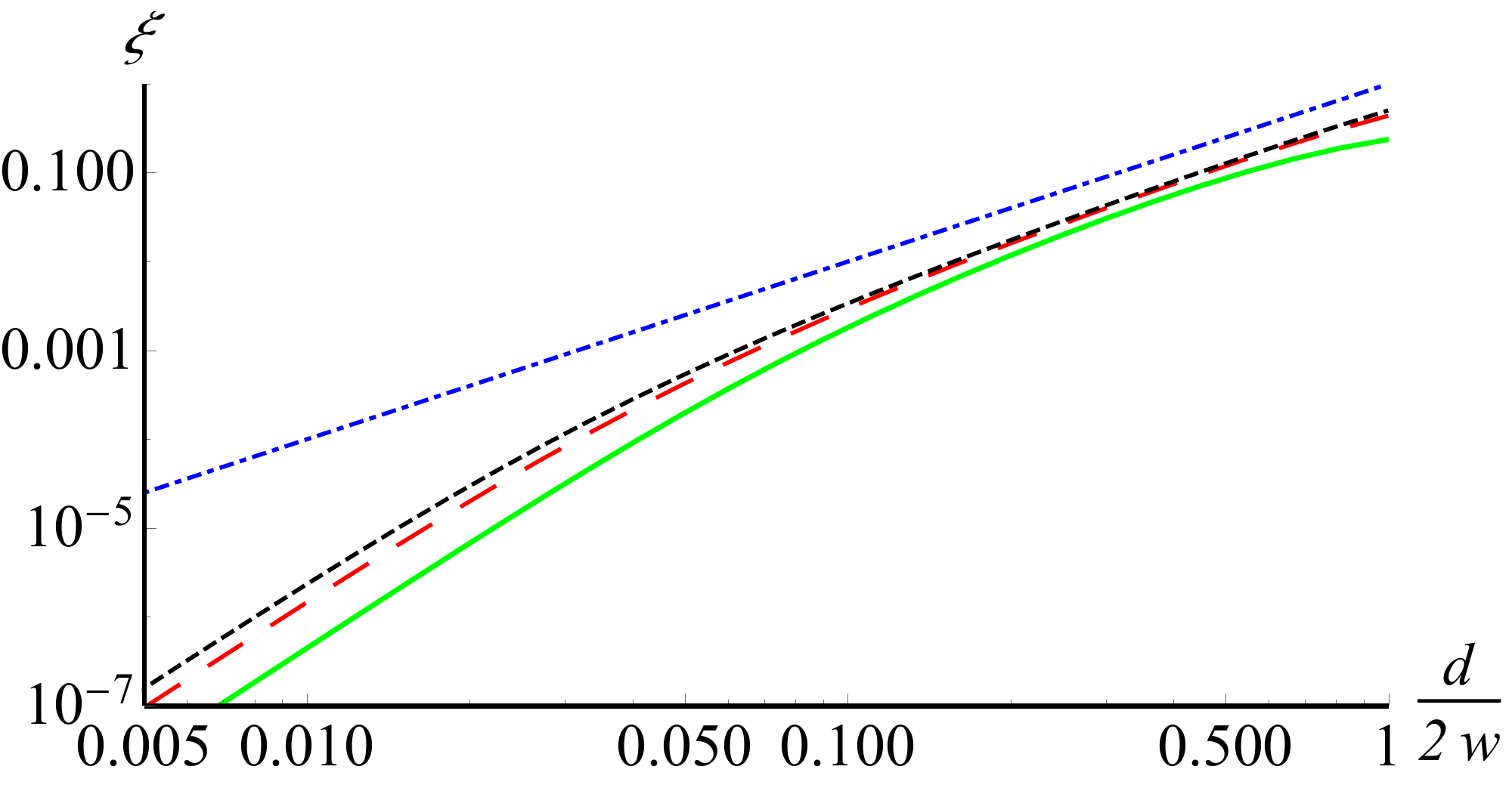}
        \caption{ Comparison of the median of Chernoff exponents for a random sample of $500$ unitary crosstalks for $D=2$ (green, continous), $D=3$ (red, long dashed),  $D=4$ (black, dashed) and quantum bound (blue, dot-dashed) versus $x$.}
        \label{fig:CHernoff2}  
\end{figure}

As explained in the main text, all our analytical analysis is performed with the assumption of $D=2$, which should be sufficient for sub-Rayleigh separations, $x= d/2w \ll 1$. To convince ourselves of that this is indeed the case, we define a new parameter, $q_c:=\epsilon^2(D^2-1)$, which stands for the average overall crosstalk probability of a single mode. In Fig. \ref{fig:CHernoff2}, we compare the median of Chernoff exponents for SPADE with random crosstalk for $D=2,3,4$ and $q_c=0.01$. 
As seen, measuring a higher number of modes for the same overall probability of crosstalk does not have a significant impact on the Chernoff exponent for small separations.

\subsection{B. Direct imaging}
For ideal, i.e. continuous and noiseless, direct imaging, the Chernoff exponent is given by
\begin{equation} \label{eq:xi_DI}
    \xi_{\textnormal{DI}}(d) =-\ln\left[ \min_{0\leq s\leq 1} \int_{\mathbf{R}^2} d\vec{r}\, p(\vec{r}\,|0)^s p(\vec{r}\,|d)^{1-s} \right],
\end{equation}
where
\begin{align} \label{eq:pdi}
    p(\vec{r}\,|d)= \frac{1}{2}\left( |u_{00}(\vec{r}-\vec{d})|^2 + |u_{00}(\vec{r}+\vec{d})|^2\right)
\end{align}
stands for the probability of measuring the photon at the location $\vec{r}$. Expanding the integrand  in $d$ and integrating in the polar coordinates, we obtain
\begin{equation} 
    \xi_{\textnormal{DI}} \approx 
        -\ln\left[ \min_{0\leq s\leq 1}1+ 4 (s-1) s x^4 \right].
\end{equation}
Minimization yields $s_{\min}=1/2$, which after another expansion to leading order in $x$ results in
\begin{equation}
    \xi_{\textnormal{DI}}\approx x^4.  
\end{equation}

\section{III. Distance independent test}\label{app:dist_ind}
\subsection{A. Optimality of the distance independent test in the ideal case}
Let us start with showing that the distance-independent test
\begin{align}
\begin{split}
    N_{10} \underset{H0}{\overset{H1}{\gtrless}} 0,\label{eq:dist_ind_ap}
\end{split}
\end{align}
is indeed equivalent to the optimal with no crosstalk present for small separations. 
Note that, since, for $x\ll 1$m the probability of photon detection for both hypotheses is mostly distributed among modes $v_{10}$ and $v_{00}$, for the total number of detected photons $N$, we have $N \approx N_{00}+N_{10}$.
Accordingly, neglecting all other modes, we can rewrite the likelihood-ratio test \eqref{eq:optimal_test} as
\begin{multline}
    p(10|x)^{N_{10}}
        [1-p(10|x)]^{N-N_{10}}   
        \\> p(10|0)^{N_{10}} [1-p(10|0)]^{N-N_{10}}.
\end{multline}
Note that, because $p(10|x)>0$, the l.h.s. is always strictly between 0 and 1. On the other hand, because $p(10|0)=0$, the r.h.s. is either equal to 0 if $N_{10}>0$, in which case H1 is accepted, or it is equal to 1 if $N_{10}=0$, in which case H1 is rejected. Clearly, this is equivalent to the distance-independent test under consideration.

However, as we discuss thoroughly in the main text, crosstalk cannot be ignored in the discussed problem. If we take crosstalk into account, can we still distinguish between the two hypotheses with a completely separation-independent test?

\subsection{B. Naive generalisation of test Eq. \eqref{eq:dist_ind_ap} in presence of crosstalk}
Let us first consider a natural generalization of the original algorithm:
\begin{align}
\begin{split}
    N_{10} &\underset{H0}{\overset{H1}{\gtrless}} N p_0.
\end{split}
\end{align}
This new algorithm is based on the fact that for a single source, i.e. H0, the expected number of photons in mode 10 equals $N p_0$, while for two sources it is (slightly) more. Note that, obviously, for $p_0\rightarrow0$ the original distant-independent test is retrieved.

However, as we will see, while this modified algorithm is better than the original one, it still has asymptotically non-vanishing probability of error, disqualifying it as a good test. To see this, we observe that in this case, both $\alpha$ and $\beta$ are given by:
\begin{align}
\begin{split}
    \alpha(N) &=  1 -F_0(N p_0), \\
    \beta(N) &= F_x(N p_0).
\end{split}
\end{align}
where $F_x(k)$ stands for the cumulative distribution function (CDF) of the binomial distribution with the probability of success $p_x$. Due to the central limit theorem, for a large number of samples the CDF of such binomial distribution is approximately equal to the CDF of the normal distribution with mean $Np_x$ and standard deviation $\sqrt{Np_x(1-p_x)}$, with equality in the limit $N\rightarrow\infty$. Using this approximation, one can easily see that:
\begin{align}
\begin{split}
    \alpha(N) &\approx 1 -\Phi\left(\frac{N p_0-Np_0}{\sqrt{Np_0(1-p_0)}}\right)=\Phi(0)=1/2,\\
    \beta(N) &\approx \Phi\left(\frac{N p_0-Np_x}{\sqrt{Np_x(1-p_x)}}\right),
\end{split}
\end{align}
where $\Phi(k)$ stands for the CDF of the standard normal distribution:
\begin{equation}
    \Phi(k)=\frac{1}{\sqrt{2\pi}}
        \int_{-\infty}^k  e^{-z^2/2}dz.
\end{equation}

Note that for small separations, we have:
\begin{equation}
    p_x\approx p_0+\gamma x^2,\label{eq:px_apr}
\end{equation}
where
\begin{equation}\label{eq:gamma}
    \gamma=|C_{10,10}|^2-p_0+\frac{2}{\sqrt{2}}\Re{C_{10,00} \left(C_{10,20}\right){}^*}
\end{equation}
is positive. Note that for $D=2$, $C_{10,20}=0$. Thus, for small separations, we have $p_x>p_0$ and therefore
\begin{equation}
    \lim_{N\rightarrow\infty} \beta(N)=\Phi(-\infty)=0.
\end{equation}
This means that the total probability of error approaches
\begin{equation}
   \lim_{N\rightarrow\infty} P_e(N)=1/4,
\end{equation}
which is far from ideal.

\subsection{C. Separation-independent tests in presence of crosstalk}
Let us now consider the more general algorithm given by Eq. (10) in the main text:\begin{align}
\begin{split}
    N_{10} \underset{H0}{\overset{H1}{\gtrless}} N p_0+\zeta(N).
\end{split}
\end{align} 
For this algorithm we analogously get:
\begin{align}
\begin{split}
    \alpha(N) &\approx 1 -\Phi_0\left(\frac{N p_0+\zeta(N)-Np_0}{\sqrt{Np_0(1-p_0)}}\right),\\
    \beta(N) &\approx \Phi\left(\frac{N p_0+\zeta(N)-Np_x}{\sqrt{Np_x(1-p_x)}}\right).
\end{split}
\end{align}
Let us consider the different possibilities for $\zeta(N)$.

\begin{itemize}
    \item If $|\zeta(N)|$ increases slower than $\sqrt{N}$ i.e.
    \begin{equation}
        \lim_{N\rightarrow\infty}|\zeta(N)|/\sqrt{N}=0
    \end{equation} 
    one can see that $\lim_{N\rightarrow\infty} \alpha(N)=1/2$.
    
    \item If $\zeta<0$ and $|\zeta(N)|$ increases faster than $\sqrt{N}$, then $\lim_{N\rightarrow\infty} \alpha(N)=1$.

    \item If $\zeta>0$ and it increases faster than $N$, then $\lim_{N\rightarrow\infty} \beta(N)=1$. 

    \item If $\zeta>0$ and it scales like $N$ i.e. $\zeta(N)=\eta N$, then if $\eta<\gamma x^2$ the probability of error vanishes. However, if $\eta>\gamma x^2$ we have $\lim_{N\rightarrow\infty} \beta(N)=1$ and for $\eta=\gamma x^2$ we have $\lim_{N\rightarrow\infty} \beta(N)=1/2$. Therefore, such test is not distance-independent, as it does not work for any $x$.
    
    \item Finally, if $\zeta>0$ and it increases faster than $\sqrt{N}$, but slower than $N$, we have $\lim_{N\rightarrow\infty} P_{\textnormal{e}}=0$ for any $x$.
\end{itemize}

\begin{figure}[t!]
    \centering
    \begin{subfigure}[b]{0.49\textwidth}
         \centering
         \includegraphics[width=\textwidth]{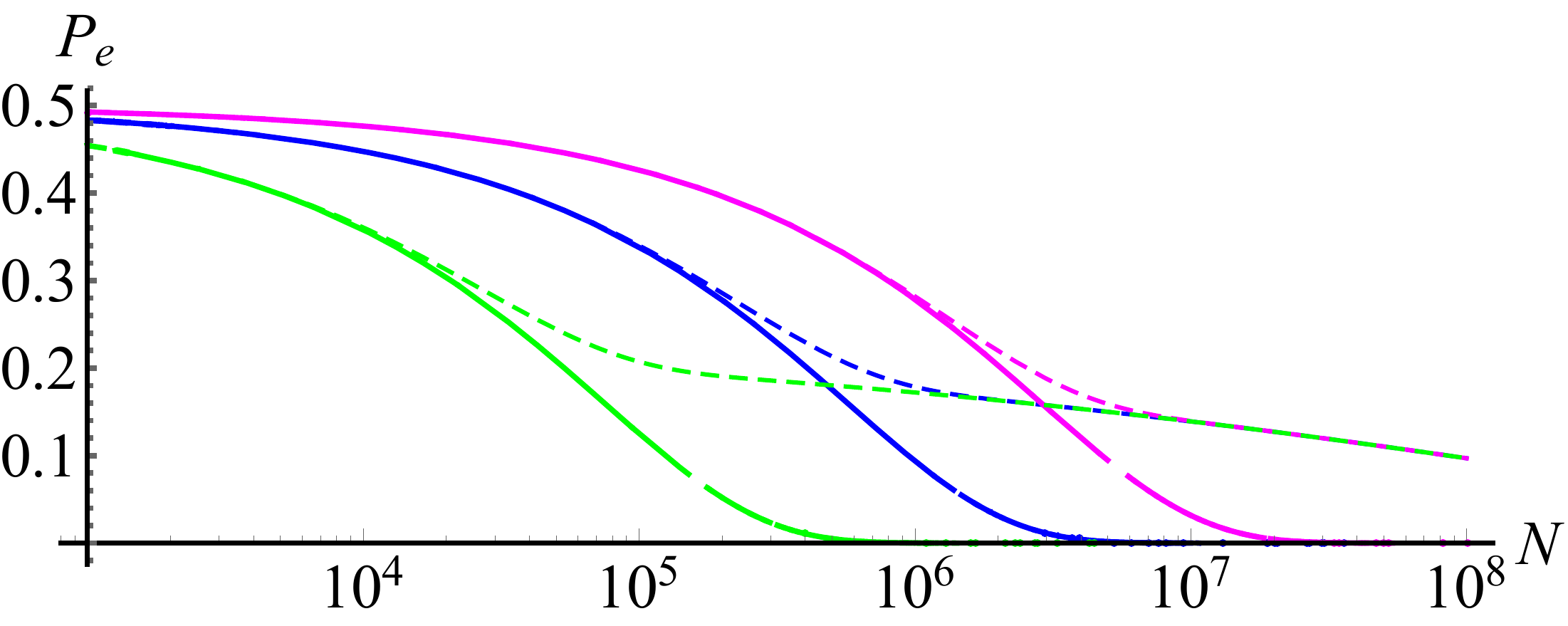}
        \subcaption[]{}
     \end{subfigure} 
      \begin{subfigure}[b]{0.49\textwidth}
         \centering
      
         \includegraphics[width=\textwidth]{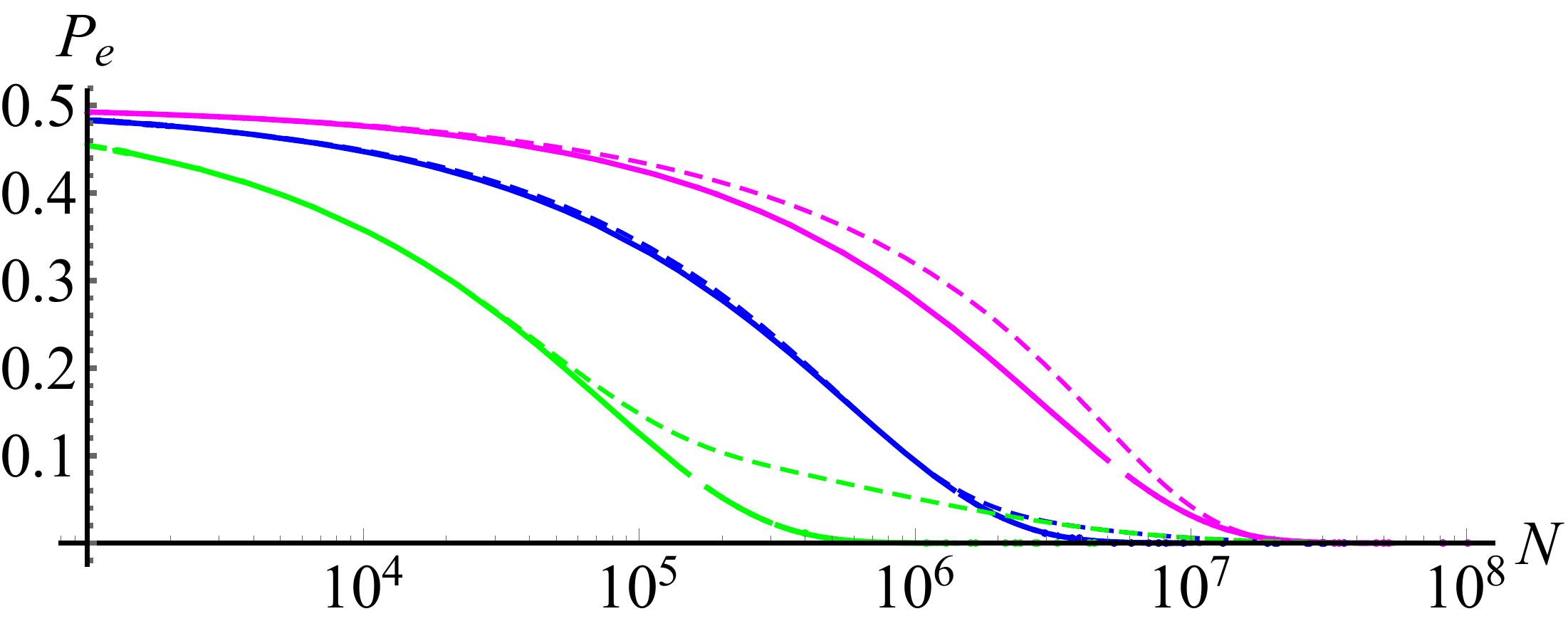}
        \subcaption[]{} 
     \end{subfigure}
      \begin{subfigure}[b]{0.49\textwidth}
         \centering
      
         \includegraphics[width=\textwidth]{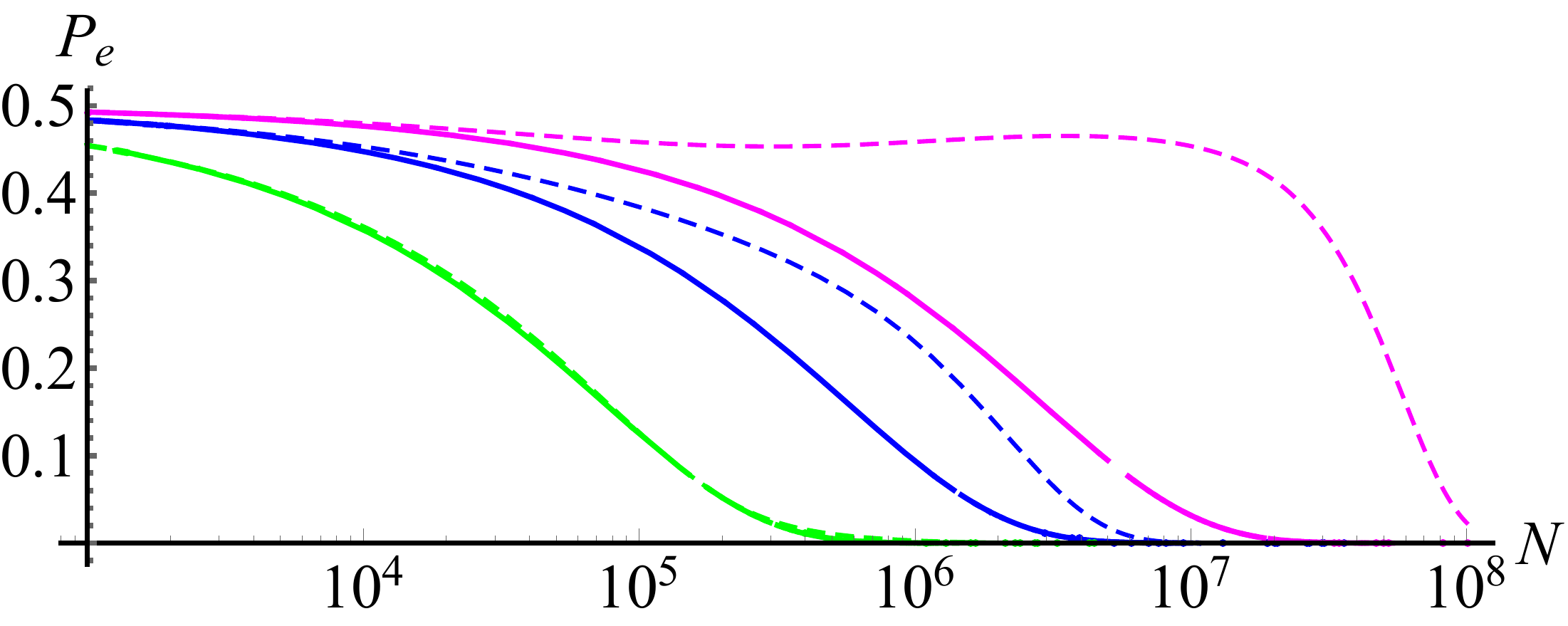}
        \subcaption[]{} 
     \end{subfigure}
        \caption{ Probabilities of error using the normal approximation to the binomial distribution versus $N$ for $\zeta=\{N^{2/3}/100,N^{3/4}/100,N^{4/5}/100\}$ (top to bottom) with uniform crosstalk for $\epsilon^2=1/15$. Magenta (top curves), blue (middle curves) and green (bottom curves) stand for for $x=\{0.02,0.03,0.05\}$, with the continuous lines corresponding to the optimal test and the dashed lines to the distance-independent tests. One can observe that for some separations, the distance-independent tests behave almost optimally, while for others they require more than an order of magnitude more samples to achieve the same probability of error as optimal test. }
        \label{fig:dist_ind}  
\end{figure}

Thus, the last case defines a class of distance-independent tests, which we discuss in the main text. 
There, we argued that this class appears to have little practical significance due to the unpredictable rates of convergence of its total probability of error.
We illustrated this behaviour for $\zeta=N^{4/5}$ in Fig.~3 a).
Analogous results for $\zeta=\{N^{2/3}/100,N^{3/4}/100,N^{4/5}/100\}$ are presented in Fig.~\ref{fig:dist_ind}.

\section{IV. Semi-separation-independent tests}\label{app:proof}
Finally, we consider the test (12) from the main text
\begin{align}
\begin{split}
    N_{10} \underset{H0}{\overset{H1}{\gtrless}} N\left(p_{0} + \gamma x^2/2\right),\label{eq:test_semi_ap}
\end{split}
\end{align}
with $x=x_{\min}$. The probabilities of error of the first and second kind for this case read:
\begin{align} \label{eq:alpha_beta_binomial}
\begin{split}
    \alpha(N) &= 1 - \sum_{k \leqslant N (p_0+\gamma x_{\min}^2/2)} 
        \binom{N}{k} p_0^k\left(1-p_0\right)^{N-k},\\
    \beta(N) &= \sum_{k \leqslant N (p_0+\gamma x_{\min}^2/2)} \binom{N}{k} p_{x_{\min}}^k\left(1-p_{x_{\min}}\right)^{N-k}.
\end{split}
\end{align} 
It is easy to see that this test resembles the case of $\zeta(N)=\eta N<N\gamma x_{\min}^2$ discussed in the previous section. Therefore, by the same arguments, the probability of error vanishes in the limit $N\rightarrow \infty$.

Now, we proceed to show that the total probability of error in the limit $N\rightarrow \infty$ vanishes also for for any separation $x>x_{\min}$ in the range where the approximation (\ref{eq:px_apr}) is valid or, in general, as long as $p_x>p_{x_{\min}}$. To see this, let us consider the change in the probabilities of error if we apply the  algorithm with $x=x_{\min}$ to hypothesis H1 with $x>x_{\min}$. By design, the threshold point does not change and it is still given by $N\left(p_{0} +\gamma x_{\min}^2/2\right)$. However, the probability of measuring a photon in mode 10 under hypothesis H1 transforms from $p_{x_{\min}}$ to $p_{x}$. Since this probability does not enter $\alpha$, this error stays the same. However, $\beta$ turns into
\begin{align}
\begin{split}
    \beta(N,x) &= \sum_{k \leqslant N (p_0+\gamma x_{\min}^2/2)} 
        \binom{N}{k} p_x^k\left(1-p_x\right)^{N-k}.
\end{split}
\end{align}

Remarkably, $\beta(N,x) < \beta(N,x_{\min})$: 
Observe that $\beta(N,x)$ is just the CDF at the point $c=\lfloor N(p_0+\gamma x_{\min}^2/2)\rfloor$, which for the case of binomial distribution can be written as
\begin{multline}
    \beta(N,x)=I_{1-p_x}(N-c,c+1)\\=(N-c)\binom{N}{c}\int_0^{1-p_x} dt\,t^{N-c-1}(1-t)^c
\end{multline}
where $I_{z}(\cdot,\cdot)$ stands for the regularized beta function. Based on this, we get:
\begin{multline} \label{eq:beta_difference}
    \beta(N,x_{\min})-\beta(N,x)=
        \\(N-c)\binom{N}{c}\int_{1-p_x}^{1-p_{x_{\min}}} dt \, t^{N-c-1}(1-t)^c
\end{multline}
where $\beta(N,x_{\min})$ is just the error of the second type for the test, in which the distance in hypothesis H1 is equal to $x_{\min}$. Because the integrand is strictly positive in the integration bounds and $p_x>p_{x_{\min}}$ [due to Eq. (\ref{eq:px_apr})], as well as the starting assumption $x\geqslant x_{\min}$, Eq. (\ref{eq:beta_difference}) is non-negative. 

\begin{figure}[t!]
    \centering
        \includegraphics[width=\columnwidth]{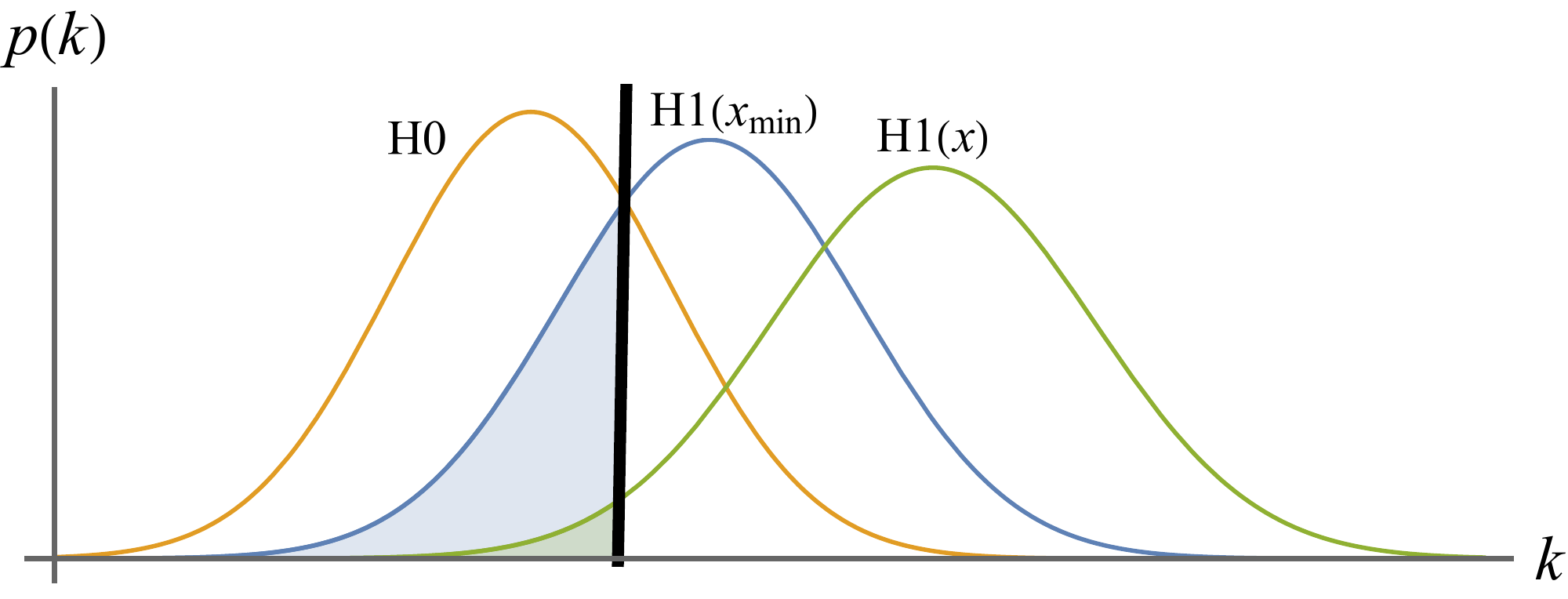}
        \caption{Probability distribution of obtaining $k$ photons in mode 10 using the Gaussian approximation to the binomial distribution. Orange (leftmost curve) stands for H0, blue (middle curve) for H1 with separation $x_{\min}$, green (rightmost curve) $H1$ for $x>x_{\min}$. Our simplified criterion is approximately equivalent to putting detection threshold at the intersection of the probability distributions for H0 and H1 with separation $x_{\min}$ (black, vertical line). The total probability of error is simply the overlapping area of these two distributions. Increasing $N$ shrinks the width of the Gaussians in relation to the separation of the centers of the peaks and therefore the overlap decreases. If $x>x_{\min}$ the center of the Gaussian is shifted towards higher $k$ (green plot) and therefore the threshold is further in the left tail of this Gaussian, resulting in a smaller contribution of such distribution towards the probability of error (green shaded area) in the semi-distance independent test. However, from this plot one can also see that such test is not optimal for the case $x>x_{\min}$. Note that considered test \eqref{eq:test_semi_ap} does not reduce to the original distance-independent test (\ref{eq:dist_ind_ap}) in the limit $p_0\rightarrow 0$, signifying a fundamental difference between testing with and without crosstalk. This qualitative difference is clearly seen in this intuitive picture, as for the case without crosstalk, the probability distribution for H0 is given by Dirac's delta $\delta(k)$, and for that reason one cannot find the point in which probability densities for both hypothesis is equal. }
        \label{fig:gauss}  
\end{figure}
To sum up, we showed that if we use the threshold point defined by the separation $x_{\min}$ for distinguishing between one or two sources at separation $x\geqslant x_{\min}$, $\alpha$ does not change, but $\beta$ becomes smaller. This means that for fixed $N$, the total probability of error for this case is actually smaller than if we had separation $x_{\min}$ between the sources:
\begin{align}
\begin{split}
    P_{\textnormal{e}}(N,x) &= \frac{1}{2}\left(\alpha(N) + \beta(N,x)\right) 
        \leqslant P_{\textnormal{e}}(N),
\end{split}
\end{align}
where $P_{\textnormal{e}}(N)=\frac{1}{2}\left(\alpha(N) + \beta(N,x_{\min})\right)$ stands for the probability off error for the test with separation in H1 equal to $x_{\min}$ in the fully distance-dependent test. 

Let us remark that the algorithm (12) can be alternatively derived by equating the binomial distributions corresponding to H0 and H1 and solving for $k$ up to the quadratic order in $x_{\min}$. This allows for an intuitive interpretation of this semi-distance independent test, as explained in Fig. \ref{fig:gauss}.

\end{document}